\documentclass[12pt]{iopart}

\usepackage[utf8]{inputenc}
\usepackage{iopams}
\expandafter\let\csname equation*\endcsname\relax
\expandafter\let\csname endequation*\endcsname\relax
\usepackage{amsmath} 
\usepackage[locale = DE]{siunitx} 
\usepackage[nointegrals]{wasysym}
\usepackage{placeins,graphicx}
\usepackage{hyperref}
\usepackage{tikz}
\usepackage[newzealand]{babel}
\usepackage[T1]{fontenc}
\usepackage{csquotes}

\definecolor{lime}{HTML}{A6CE39}
\newcommand{\orcidicon}{%
	\begin{tikzpicture}
	\draw[lime, fill=lime] (0,0) 
	circle [radius=0.16] 
	node[white] {{\fontfamily{qag}\selectfont \tiny ID}};
	\draw[white, fill=white] (-0.0625,0.095) 
	circle [radius=0.007];
	\end{tikzpicture}
	\hspace{-3mm}
}
\newcommand\orcidMatt{{\href{https://orcid.org/0000-0003-1088-6485}{\orcidicon}}}
\newcommand\orcidSeSc{{\href{https://orcid.org/0000-0003-1997-0026}{\orcidicon}}}

\newcommand{\kl}[1]{\left( #1 \right)}
\newcommand{\kle}[1]{\left[ #1 \right]}
\newcommand{\ed}[1]{\frac{1}{#1}}
\newcommand{\defi}{\mathrel{\mathop:}=}
\newcommand{\ifed}{=\mathrel{\mathop:}}
\newcommand{\pdet}{\ensuremath{\mathrm{pdet}}} 
\newcommand{\dif}{\ensuremath{\mathrm{d}}}
\newcommand{\eps}{\varepsilon}
\newcommand{\mi}{\ensuremath{{\left[\mu^{-1}\right]}}}
\newcommand{\gi}{\kle{g_{\text{eff}}^{-1}}}
\newcommand{\g}{g_{\text{eff}}}

\usepackage[sorting=none,doi=true,eprint=true,arxiv=abs,hyperref=true,bibstyle=nature-modified,citestyle=numeric]{biblatex}
\DeclareNameAlias{default}{first-last}

\begin{document}
	\title[Electromagnetic analogue space-times]{Electromagnetic analogue space-times, analytically and algebraically}
	\author{Sebastian Schuster\orcidSeSc and Matt Visser\orcidMatt}
	\address{School of Mathematics and Statistics, Victoria University of Wellington, \\
		\qquad PO Box 600, Wellington 6140, New Zealand.}
	\eads{\mailto{sebastian.schuster@sms.vuw.ac.nz}, \mailto{matt.visser@sms.vuw.ac.nz}}
	\begin{abstract}
		While quantum field theory could more aptly be called the \enquote{quantum field framework} --- as it encompasses a vast variety of varying concepts and theories --- in comparison, relativity, both special and general, is more commonly portrayed as less of a \enquote{general framework}. Viewed from this perspective, the paradigm of analogue space-times (also often called analogue gravity) is to promote the specific theory of general relativity (Einstein gravity) to a \emph{framework} which covers relativistic phenomena at large. Ultimately, this then also gives rise to new proposals for experiments in the laboratory, as it allows one to move general features of the \enquote{relativistic framework} from general relativity to entirely new new areas of physics. This allows one to experimentally look into analogies of currently unobservable phenomena of general relativity proper. The only requirement for this to work is the presence of a notion of an upper limit for propagation speeds in this new setting. Systems of such a kind abound in physics, as all hyperbolic wave equations fulfil this requirement.
		
		Consequently, models for analogue space-times can be found aplenty. We shall demonstrate this here in two separate analogue space-time models, both taken from electrodynamics in continuous media. First of all, one can distinguish between analytic analogue models (where the analogue is based on some specific hyperbolic differential equation), on the one hand, and algebraic models (where the analogue is fashioned from the more or less explicit appearance of a metric tensor), on the other hand. Yet this distinction is more than just a matter of taste: The nature of the analogue space-time model will also determine which physical concepts from general relativity can be taken easily into an experimental context. Examples of this will constitute the main aim of this paper, and the Hawking effect in one of the two models considered the example of most immediate experimental interest.
	\end{abstract}
	
	
	\maketitle
	
	\section{Introduction: Analogue Space-Times and Electromagnetism in Continuous Media}
	Before starting the discussion proper, it seems useful to remind ourselves of the currently favoured take on \enquote{analogue space-times}. The reasoning in this case is based on our understanding of general relativity (and its offspring with quantum field theory, curved space-time quantum field theory) in its natural habitat: astronomy, astrophysics, and cosmology. Many of the effects and features most interesting to the theorist at large are very far removed from experimental accessibility: Gravitational waves (only recently, and famously so, observed; albeit after a colourful history), horizons (slowly becoming observable thanks to the efforts of the Event Horizon Telescope), inflation (with a chequered experimental history to date), the Hawking effect (being central driver of the analogue space-time paradigm and seeing first successes), and many further minute effects beyond the standard model of either cosmology or particle physics. Were one now to follow the spirit of relativity in a way more seen as a \emph{framework} for a great many different theories (beyond its astrophysical origin) --- as demonstrated in \cite{EssIness} for quantum effects in curved space-times --- new physical effects can be encountered, both entirely unknown to general relativity or analogues of general relativity's effects. The hope is that one will be able to identify many ideas originally found in relativity in new contexts --- contexts which are far more amenable to experimental laboratory success. Recent developments in laboratories show that this hope is not misplaced \cite{BECHawkingExp2018,SuperradianceDetected,RousseauxExp2016,SteinhauerExp2,SteinhauerExp1,UnruhHawkingMeasured,MeasurementWhiteHoleHawking}.
	
	It is worth noting that this (somewhat) operationalist interpretation is not the original impetus behind the analysis: The developments usually came from two different approaches. The first was the matter-of-fact theoretical observation of the similarities, for example in Gordon's article \cite{GordonMetric}, while the second had aid to gravitational physics from the experience available for different fields of physics (in the present instance: electromagnetism) in mind. In the words of de~Felice:
	\blockcquote{deFeliceTrafoOptics}{[\dots] it may be mathematically more cumbersome and less elegant. However, owing to the more familiar and physically more intuitive type of approach the equivalence which is here proposed might help whenever one wants to solve new problems in gravitation, such as that of physical optics.}
	More on the history relevant to the present analysis can be found in section~3.1.1 of~\cite{lrrAnalogue} and de~Felice's article~\cite{deFeliceTrafoOptics}.
	
	These differences in approaching the subject certainly do not change the physical content of the calculations to be presented. Being aware, however, of the specific approach taken here will help appreciating our conclusions. This is especially true for the \textit{first example} to be discussed: An \emph{algebraic} analogue of vacuum electrodynamics in a specified metric $\g$ and electrodynamics in a continuous medium (in some arbitrary background metric $g$). The reason for us calling it \enquote{algebraic} relates to the fact that after a close examination of the analogue, its origin relies only on an algebraic combination of tensors. We will find in the next section that matters put to rest in the usual context of astrophysical general relativity, particularly those pertaining to coordinate independence, here gain a new \textit{physical} aspect. 
	
	More on the context in which this algebraic analogue space-time is encountered can be found in \cite{lrrAnalogue}, \cite{CovEffMetrics}, and references therein. In our article \cite{CovEffMetrics} we collected, unified and generalised previously known results for this analogue. (Specifically, we were interested in expressing the effective metric in terms of the material properties; provided the material satisfied the necessary and sufficient consistency conditions for representing an analogue space-time model.) A similar, modern analysis of macroscopic electrodynamics has been undertaken in the field of (transformation) optics, to be found in \cite{PerlickRay,BalZim,CovOptMet1,CovOptMet2,CovOptMet3,TrafoOpticsCartographDistort,CovOptMet4}. The unifying point of all these developments was a notion of truly four-dimensional formalisms, while the pre-metric approach to electrodynamics (see, for example, \cite{HehlObukhov,HehlLaemmerzahl,FavaroBergamin,HehlAxionDilaton}) takes an orthogonal approach starting from a $3+1$-dimensional viewpoint to derive the four-dimensional metric. Similar to our algebraic analogue, teleparallel gravity has been identified as an analogue for non-linear electordynamics \cite{TeleparallelNonlinEdynAnalogue}, though from a more theoretical vantage point with less experimental implications in mind.
	
	The \textit{second example} will be slightly less plagued by hazards hidden in the physical interpretation of mathematical symbols. It mimics aspects of the wave equation on a curved space-time through the use of refractive indices and their appearance in the Helmholtz equation of a medium. While the origin of this example is of less theoretical import than that of the algebraic analogue, we imagine it to be experimentally easier to implement, both with regard to the interpretation and to the technical considerations. From a theoretical point it serves to highlight the fact that analogue models can look for other structures than an effective metric $\g$ --- here we will see that the analogy lies entirely in the realm of separated wave equations without (directly) invoking the metric. Its origin in the realm of differential equations explains why we call it an \emph{analytic}. As its scope is more limited than the full analogue space-time paradigm, we will mostly refer to it as an \enquote{analogy} rather than an \enquote{analogue}. This minor linguistic quibble aside, this analogy will serve as a good example to demonstrate our desire to distinguish \enquote{algebraic} from \enquote{analytic} analogies/analogues, as our title tries to emphasize.
	
	Finally, we will close by collecting our results, and suggesting possible extensions.
	
	\subsection*{Notation}
	Before beginning, let us fix our notational choices: Our (Lorentzian) metrics will be of signature $(-+++)$. General space-time indices will be labelled with Latin letters starting from $a, b, \dots$. Where only index placement, but not explicit labelling is required, the corresponding space-time index will be denoted by the symbol $\bullet$. Purely spatial indices will be labelled with Latin letters starting from $i, j, \dots$. Where only index placement, but not explicit labelling is required, the corresponding space index will be denoted by the symbol $\circ$. Levi-Civita (pseudo-)tensors are denoted by $\eps$, while Levi-Civita tensor densities are written using $\tilde \eps$. Two metrics, $g$ and $\g$, are used throughout the discussion; raising and lowering of indices will \emph{always} be done with $g$. We will mostly use natural units, and make note of the exceptions where the discussion necessitates them.
	
	\section{An Algebraic Analogue}\label{sec:alg}
	
	In our algebraic analogue, the ingredients are two metrics (the background or laboratory metric $g$, and the effective or analogue metric $\g$), and an electromagnetic medium with permittivity $\epsilon^{ij}$, permeability $\mi^{ij}$, and magneto-electric effect $\zeta^{ij}$. The matrices $\epsilon$ and $\mi$ are real-valued, symmetric $3\times 3$ matrices, while $\zeta$ is a general (possibly asymmetric, possibly vanishing), real-valued $3\times 3$ matrix. Were we to allow complex values, this would lead to dissipation and break the Lorentzian analogy. As we will see below, this would lead to complex entries in the effective metric $\g$ --- for our purposes a rather unsavoury feature. We shall refer to the set of these three matrices as the \enquote{constitutive matrices}. The first step is to translate this into a truly four-dimensional formalism.
	
	\subsection{The Formalism}
	In order to achieve this four-dimensional notation, one supplements the field strength tensor $F^{ab}$ with the excitation tensor
	\begin{equation}\label{eq:constitutive}
	G^{ab} \defi Z^{abcd}\,F_{cd}.
	\end{equation}
	Here, $Z$ is a fourth rank tensor --- the \enquote{constitutive tensor} --- with the following properties:
	\begin{equation}
	Z^{abcd} = Z^{cdab}, \qquad Z^{(ab)cd} = Z^{ab(cd)} = 0.\label{eq:Zsym}
	\end{equation}
	This tensor encodes the material properties previously contained in the three $3\times 3$ matrices $\epsilon, \mu^{-1}$, and $\zeta$. It is their four-dimensional generalisation, like the field strength tensor is the four-dimensional generalisation of magnetic flux $\mathbf{B}$ and electric field $\mathbf{E}$. In the context of this paper, it is useful to make this more explicit: Any two-form (or two-vector) in 3+1 dimensions can be written using one four-velocity and two associated one-forms (or vectors). For $F$, this means that it can be written in terms of an observer's notion of electric and magnetic fields, $E$ and $B$, respectively, and their four-velocity $V$, 
	\begin{equation}\label{eq:Fdecom}
	F_{ab} = V_a E_b - V_b E_a + \eps_{abcd} V^c B^d.
	\end{equation}
	This can be achieved for the excitation tensor $G$, too. However, in this case, the relevant fields are the electric and magnetic excitation, $D$ and $H$, fulfilling
	\begin{equation}\label{eq:Gdecom}
	G^{ab} = V^b D^a - V^a D^b + \eps^{abcd} V_c H_d.
	\end{equation}
	$E$, $B$, $D$, and $H$ are all four-orthogonal to $V$. This corresponds to the familiar $3+1$ version of the constitutive relations
	\begin{subequations}
		\begin{alignat}{4}
		\mathbf{D} &=& \epsilon\; \mathbf{E}\; &+& \zeta\; \mathbf{B},\\
		\mathbf{H} &=& \zeta^\dagger \mathbf{E}\; &+&\; \mu^{-1} \mathbf{B}.\label{eq:EBtoDH}
		\end{alignat}
	\end{subequations}
	
	It is possible to consider less restrictive constitutive tensors (for example, this is often encountered in a pre-metric context as done in \cite{HehlObukhov}), or considering a minor variation of the constitutive relations which gives rise to slightly different magneto-electric tensors, see \cite{AniBiGuide,TrafoOpticsCartographDistort} --- concretely, this is the difference between Tellegen and Boys--Post constitutive relations. We will be using the latter. This is, again, a matter of convention rather than of physics. Another option is to use more restrictive constitutive tensors (by demanding the first (algebraic!) Bianchi identity, $Z^{[abcd]}=0$, to hold, as done in \cite{Post}, corresponding to a vanishing axion field \cite{HehlObukhov}). Our later application of a constitutive tensor to describe an analogue space-time will actually result in the first Bianchi identity to be fulfilled. The Bianchi identity is connected to the vanishing of the fully antisymmetric part of $Z$. This vanishes as in our case (yet to be described) the ingredients are symmetric second rank tensors, the inverse effective metrics. But at this stage it is not required to demand its validity, and we shall therefore refrain from doing so. Depending on one's choice of definition of constitutive tensor, $G$ may have a similar property as the dual field strength tensor $\star_g F$ (where $\star_g$ denotes the Hodge star operator of metric $g$): The positioning of electric and magnetic parts may change compared to that in $F$ itself: While the electric fields $\mathbf{B}$ are the purely spatial part of the field strength tensor, in $\star_g F$ (or possibly $G$) this role will be filled by $\mathbf{D}$, similarly for $\mathbf{E}$ and $\mathbf{H}$ in the spatio-temporal part of $F$ and $\star_g F$ (or possibly $G$). Our definition of $G$ does not do this --- the spatio-temporal components are $\mathbf{D}$, the purely spatial ones are $\mathbf{H}$.
	
	It is now possible to write down four-dimensional expressions for our above-introduced constitutive matrices: 
	\begin{subequations}\label{eq:CovConstMatrices}
		\begin{align}
		\epsilon^{ab} &\defi - 2 Z^{acbd} V_c V_d, \label{eq:defeps4}\\
		\mi^{ab} &\defi \frac{1}{2} \eps^{ca}{}_{ef}\eps^{db}{}_{gh} Z^{efgh} V_c V_d, \label{eq:defmu4}\\
		\zeta^{ab} &\defi \eps^{ca}{}_{ef} Z^{efbd} V_c V_d. \label{eq:defZeta4}
		\end{align}
	\end{subequations}
	It is easy to check that this agrees with our experience from the case of vacuum electrodynamics (\emph{i.e.}, microscopic electrodynamics) where
	\begin{equation}\label{eq:Zvacuum}
	Z^{abcd}_\text{vacuum} \defi \ed{2}\kl{g^{ac} g^{bd} - g^{ad}g^{bc}},
	\end{equation}
	which results in the familiar action
	\begin{subequations}
		\begin{align}\label{eq:Svacuum}
		S_\text{vacuum} &= -\ed{4} \int \dif^4x \sqrt{\det g}\; F_{ab} F^{ab},\\
		&= -\ed{8} \int \dif^4x \sqrt{-\det g} \kl{g^{ac}g^{bd} - g^{ad}g^{bc}}F_{ab}F_{cd}.
		\end{align}
	\end{subequations}
	
	Let us now give the corresponding, macroscopic Maxwell equations:
	\begin{subequations}\label{eq:Maxwell}
		\begin{align}
		\nabla_a G^{ab} &= J^b, \label{eq:InHomMaxwell}\\
		\nabla_{[a} F_{bc]} &= 0.\label{eq:HomMaxwell}
		\end{align}
	\end{subequations}
	While not necessarily needed for the analogue space-time itself, these will later play an important role in our discussion of the meaning of coordinate invariance for the analogue space-time.
	
	\subsection{The Analogy}
	The analogy is now made by comparing a naive constitutive tensor for vacuum electrodynamics with respect to a chosen and fixed \enquote{target metric} $\g$,
	\begin{equation}
	Z^{abcd} = \ed{2}\kl{\gi^{ac}\gi^{bd} - \gi^{ad}\gi^{bc}},\label{eq:ZeffPattern}
	\end{equation}
	with that of a meta-material in a laboratory (with some background laboratory metric $g$). In preparation for what is to come we already made the distinction between the fully contravariant metric and the inverse metric explicit. At this level of the discussion this is not yet necessary, but it will make the transition to what is needed in the end smoother. The idea behind equation~\eqref{eq:ZeffPattern} lies in the fact that it corresponds to what we know from vacuum electromagnetism, as seen in equation~\eqref{eq:Zvacuum}. However, now it is not an actual vacuum space-time giving the metric. Rather, we have a medium (with electromagnetic properties encoded in $Z$) mimicking the \enquote{constitutive relations} of a different vacuum space-time with metric $\g$.\footnote{Note that in transformation optics, particularly in the paper~\cite{CovOptMet3}, a careful analysis already has been undertaken to make this statement more rigorous. To achieve this, the influences of background metric and material properties are carefully identified and separated. While we agree with the general results, it should be pointed out that equation~(31) of \cite{CovOptMet3} enforces an isometry between effective and background metric. This is not necessary, and a too strong restriction.}
	
	It is now important to note that all that happened is a purely algebraic process: We constructed explicitly a constitutive tensor $Z$ in such a way that the corresponding, macroscopic Lagrangian at \emph{first} glance looks like the Lagrangian of vacuum electrodynamics of a manifold with metric $\g$. As we shall see below, this will turn out to not be the case.
	
	In order to arrive at the right dynamics, that is, the correct effective vacuum Maxwell equation as they appear in equations~\eqref{eq:Maxwell}, we now need to have a closer look at the corresponding action functional. What we \emph{want} is an action that reads
	\begin{equation}
	S = -\ed{8} \int \dif^4x \sqrt{-\det \g} \kl{\gi^{ac}\gi^{bd} - \gi^{ad}\gi^{bc}}F_{ab}F_{cd}.\label{eq:actioneff}
	\end{equation}
	It is important to now note that this differs from the action the laboratory would prefer:
	\begin{align}
	S &=-\ed{4}\int \dif^4x \sqrt{-\det g}\; Z^{abcd}F_{ab} F_{cd},\label{eq:actionefflab1}\\
	&=-\ed{8} \int \dif^4x \sqrt{-\det g} \kl{\gi^{ac}\gi^{bd} - \gi^{ad}\gi^{bc}}F_{ab}F_{cd}.\label{eq:actionefflab2}
	\end{align}
	It is easy to miss that this difference in the volume element of equation~\eqref{eq:actionefflab1} and equation~\eqref{eq:actionefflab2} results in different equations of motion. To still satisfy the two requirements
	\begin{itemize}
		\item that the correct effective vacuum electrodynamics arise out of the action~\eqref{eq:actionefflab1},
		\item and that $Z^{abcd}$ transforms as a tensor. (Admittedly, this is a matter of taste. It is possible to use densities. We opted for the fully tensorial approach.)
	\end{itemize}
	the tensor
	\begin{equation}
	Z^{abcd} = \ed{2}\frac{\sqrt{\det \g}}{\sqrt{\det  g}} \kl{\gi^{ac}\gi^{bd} - \gi^{ad}\gi^{bc}}\label{eq:Zeff}
	\end{equation}
	might be employed. Note that the presence of \emph{two} metrics (background and effective) cannot be captured by the naive constitutive tensor~\eqref{eq:ZeffPattern} as found in microscopic electrodynamics. With this more elaborate constitutive tensor~\eqref{eq:Zeff}, it might now be possible to describe the electrodynamics in the medium in the laboratory in terms of the vacuum electrodynamics with a given effective metric $\g$.
	
	That also this hope is in vain, can be most clearly seen from the fact that both micro- and macroscopic electrodynamics in four space-time dimensions is conformally invariant. As the physical metric is the laboratory metric $g$, we look at conformal transformations of it, not of $\g$. The additional factor ${\sqrt{\det \g}}/{\sqrt{\det  g}}$ thus has to change the dynamics, as the Lagrangian \emph{density} (which includes $\sqrt{\det g}$) should have conformal weight 0: For macroscopic electrodynamics in $D+1$ space-time dimensions, $F$ has weight 0, $\sqrt{\det g}$ weight $D+1$, and $Z$ weight $-D-1$. Coincidentally, this reproduces the result that microscopic electrodynamics is conformally invariant only in $D=3$; easily verified by looking at equation~\eqref{eq:Svacuum}.
	
	Here we can also see the need to distinguish between the twice contravariant effective metric and the inverse effective metric --- with respect to the background metric these will be different tensor fields, as it is the background metric that will define metric duality, \emph{i.e.}, which raises and lowers indices. The tensor $\g^{ab} = [\g]_{cd}\, g^{ac} g^{bd}$ thus is not the inverse to $[\g]_{ab}$, by definition $\gi^{ab}$ is. Contrast this with $[g^{-1}]^{ab} = g^{ab}$.
	
	\subsection{Bespoke Meta-Material Mimics and Coordinates: Cartographic Distortions}\label{sec:coords}
	As a simple degrees of freedom (d.o.f.) counting shows, any metric one might want to mimic (10~d.o.f.) can be captured by a choice of appropriate material properties $Z$ (21~d.o.f.). All one has to do is choose the metric $\g$ one is interested in, and use either equations~\eqref{eq:CovConstMatrices} or the consistency conditions resulting from the mismatch of d.o.f. to calculate the constitutive tensors for this metric. In two sequels to \cite{CovEffMetrics}, namely \cite{Bespoke1} and \cite{Bespoke2}, we gave explicit examples of constitutive tensors corresponding to several standard forms of black hole space-times $\g$. Already at this level, it becomes noticeable that a simple difference in the coordinates chosen to represent $\g$ will drastically change the electromagnetic properties of a laboratory material supposed to mimic this metric. The quickest way to realise this is by looking at the defining equation~\eqref{eq:defZeta4} for the magneto-electric tensor, $\zeta$: Take the Kerr metric in Boyer--Lindquist coordinates for $\g$,
	\begin{equation}
	g^\text{eff}_{ab} = \left(\begin{array}{c|cc|c} 
	g^\text{eff}_{tt}  & 0 & 0& g^\text{eff}_{t\phi}\\
	\hline
	0 & g^\text{eff}_{rr} &0 &0\\
	0&0&g^\text{eff}_{\theta\theta}&0\\
	\hline
	g^\text{eff}_{\phi t}&0&0&g^\text{eff}_{\phi\phi}
	\end{array}\right).\label{eq:BoyerLindquist}
	\end{equation} 
	The matrix representation of $\gi$ is then not diagonal, hence $\zeta$ will not vanish. On the other hand, we know that (at least in sufficiently small, topologically trivial coordinate patches of the domain of outer communication, away from the ergo surface) we can choose a coordinate system which diagonalises $\g$. Hence, in these new coordinates, $\zeta$ \emph{does} vanish. And this can be achieved long before introducing the more general coordinate transformations usually encountered in a relativistic context.\footnote{It is worth pointing out, however, that the linear transformation used in the above argument will mix expression involving coordinates in rather haphazard ways. In calculations this will rather obfuscate than clarify.}
	
	At this point, confusion might arise: The physics is the same, so why do the electric properties of the material mimicking them change? Obviously, different material properties have different physical implications in the laboratory. At the very least, a material with non-vanishing magneto-electric effects will behave non-locally, see \cite{Post}, unlike one where $\zeta=0$.\footnote{This is \emph{not} a violation of causality. It is just an effect of the averaging performed when passing from the microscopic (and fully local!) electrodynamics to the macroscopic electrodynamics. Chiral molecules (for example) will break the locality of the resulting macroscopic theory, while being microscopically perfectly local.}
	
\begin{figure}
		\includegraphics[width=\textwidth]{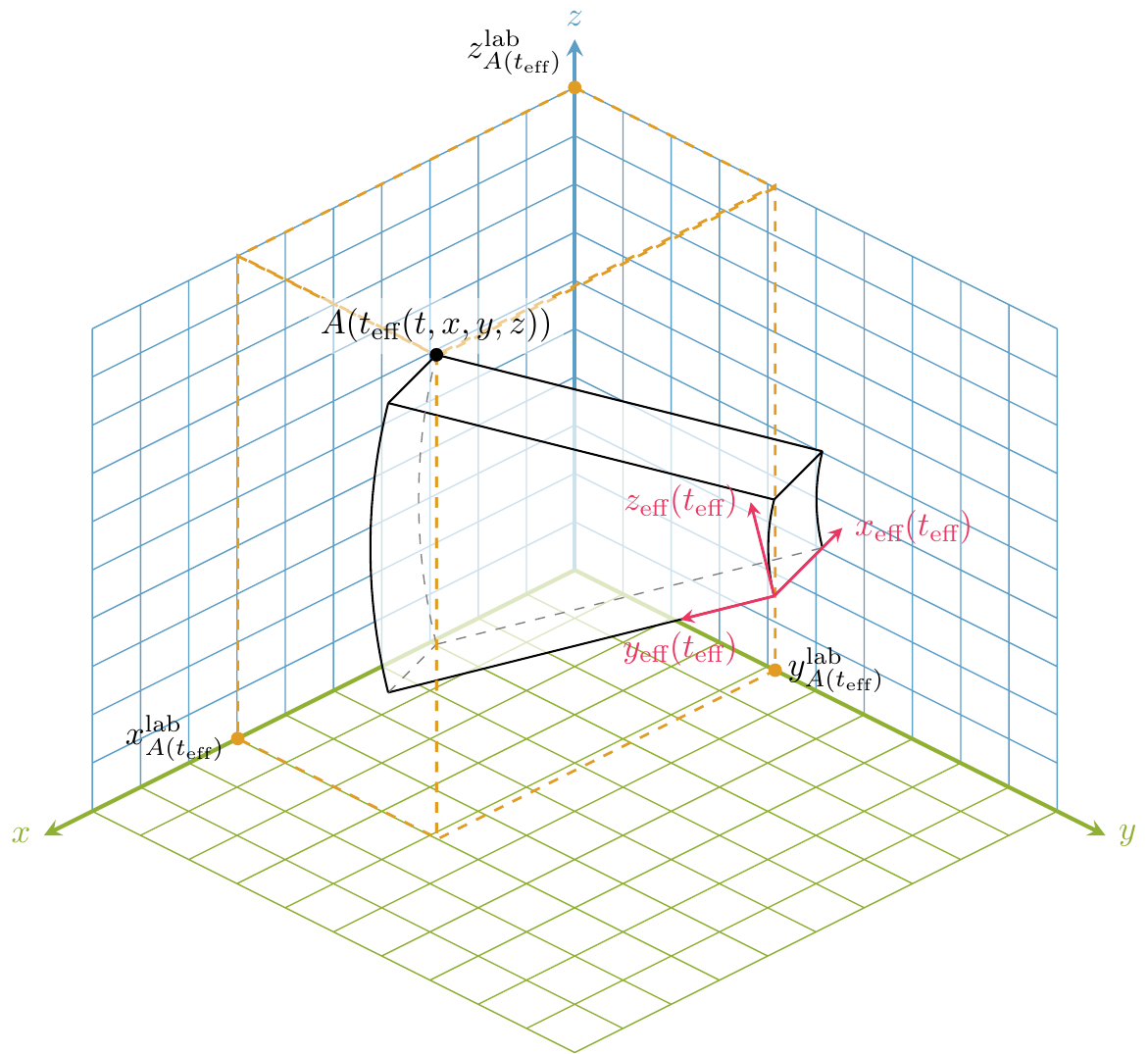}
		\caption{A sketch of the difference between analogue and laboratory coordinates. Note that every instance of $t_\text{eff}$ will depend on the laboratory coordinates $t,x,y,z$. The meaning of a given laboratory coordinate can change with respect to the effective space-times time coordinate.}
		\label{fig:analoguecoords}
	\end{figure}

	In order to illuminate this situation, let us look at the situation from a slightly more abstract point of view: A \enquote{space-time} is a Lorentzian manifold $(M,g)$, that is, a (smooth) manifold $M$ and a Lorentzian metric $g$ defined on it. The experimenters and their lab are in the physical space-time $(M,g)$. However, a part of this space-time, say $U\subset M$ is taken up by the material mimicking our target metric $\g$. The effective Lorentzian manifold then is $(U,\g)$. Obviously, the space-times are not the same, not even $(U,g)$ and $(U,\g)$.\footnote{Unless, for whatever reason, one would want to mimic the background metric itself, for which a vacuum is then perfectly sufficient. No complicated analogue space-time analysis will be necessary. Let us ignore this rather superfluous scenario.} (Topologically, $(U,g)$ and $(U,\g)$ \emph{are} the same, and $U$ is a submanifold of $M$.) Nonetheless, one can still use the coordinates $t,x,y,z$ of the background manifold to label the coordinates $t_\text{eff}, x_\text{eff}, y_\text{eff}, z_\text{eff}$ of the effective manifold. The physics as seen within the effective space-time $(U,\g)$ \emph{do not} change under a different coordinate system for $(U,\g)$. The labelling in the laboratory, however, will --- and so do the requirements on the material used to capture this labelling, hence in turn also the physics used in the laboratory to mimic the effective space-time. An attempt at picturing this relation between background coordinates and effective coordinates is shown in figure~\ref{fig:analoguecoords}. This also explains our mentioning \enquote{cartography} in this section's title. It is precisely the situation encountered in classical cartography: A curved surface is depicted on the flat surface of a page in an atlas. Our labelling of the effective space-time's coordinates with the laboratory space-time's coordinates corresponds exactly to cartographic projections --- the only difference being a doubling of the dimensions involved, $4$ instead of $2$.
	
	At first, this might look ominous. We think rather that this opens up a lot of options for mimicking any given space-time: Should the requirements on a given (meta-)material be too unrealistic to be experimentally available, one can look for other coordinates in which a mimicking material can be found. A different kind of experimental complication, however, will still arise.
	
	In the context of transformation optics, Fathi and Thompson have described this under the very fitting moniker \enquote{cartographic distortions} \cite{TrafoOpticsCartographDistort}. In the next two sections, we will look at this more closely. This caveat known from transformation optics is also a good opportunity to mention the experimental situation: Transformation optics usually has slightly different goals than the analogue space-time paradigm. This notwithstanding, transformation optics' claim to be able to provide cloaking devices certainly boosted its experimental realisation \cite{TrafoOptics2,CovOptMet2}. This is done through the help of meta-materials, whose \enquote{cloaking} properties are constrained to a very narrow microwave bandwidth. Similar devices using similar meta-materials can thus easily be used for (similarly limited bandwidth) application within the analogue space-time paradigm.
	
	As a last remark along these lines, it can certainly be said that given the non-tensorial nature of part of de~Felice's analysis \cite{deFeliceTrafoOptics} his results already anticipated these cartographic effects.
	
	\subsection{Complications Related to the Maxwell Equations}
	A different kind of trouble is less directly related to the coordinates: In this case, the issue is with the differing metrics. The effective vacuum electrodynamics, as encapsulated in the equations~\ref{eq:Maxwell}, are only valid within the framework of the effective space-time $(U,\g)$. In particular, this means that our covariant derivative appearing in the Maxwell equations will be unrelated to the covariant derivative used by experimenters in $(M,g)$. (Should one prefer one's Maxwell equations written in terms of forms, the issue is again encapsulated in the metric dependence of the Hodge star operator as well as in the necessity to lower the indices of $G^{ab}$ with the background metric $g$ to have a well-defined form. Defining the constitutive tensor slightly differently and more fitting for forms as $\tilde{Z}_{ab}{}^{cd}$ would just move this to a matter of raising the indices of the form $\tilde{G}_{ab} = \tilde{Z}_{ab}{}^{cd}F_{cd}$. For reasons to be mentioned soon, we prefer using covariant derivatives.) 
	
	We will see that this is where the simple \emph{algebraic} analogy complicates matters. In many analogue space-time models, see for example those considered and referenced in \cite{lrrAnalogue}, the starting point is a wave equation in whose structure an effective metric is recovered --- hence our calling this type of analogue \emph{analytic}. Here, however, the analogy did not start from a wave equation (or, in the absence of a four-potential to make the wave equation from the Maxwell equations explicit, the Maxwell equations themselves) --- the analogy was purely on an algebraic level inspired by the raising and lowering of indices as encoded in $Z^{abcd}_\text{vacuum}$.
	
	\noindent To simplify the discussion, we shall make two assumptions in this subsection:
	\begin{enumerate}
		\item The labelling $(M,g)\to(U,\g)$ is the identity map.
		\item We impose the topological constraints necessary for having a four-potential $A_a$ available.
	\end{enumerate}
	Using the second assumption, we can then write the inhomogeneous Maxwell equation in the form of a wave equation:
	\begin{equation}
	\nabla_{a}^{\text{eff}} \kl{\nabla^{a}_\text{eff} A^b - \nabla^{b}_\text{eff} A^a} = J^b_\text{eff},\label{eq:wave}
	\end{equation}
	or more explicitly,
	\begin{equation}
	\ed{\sqrt{\det\g}}\partial_a \kl{\sqrt{\det\g} \gi^{ad} \gi^{cb} \kl{\partial_d A_c - \partial_c A_d}}= J^b_\text{eff}.\label{eq:wave2}
	\end{equation}
	The issue becomes obvious once one reminds oneself that $\nabla^\text{eff}$ is the covariant derivative with respect to the \emph{effective} Levi-Civita connection only. This is \emph{not} what the laboratory will be using in its measurements. The experimenters' electrodynamics is governed by the Levi-Civita connection of the \emph{background} metric. The electromagnetism of an actual, physical space-time with metric $\g$ would raise the indices in equation~\eqref{eq:wave} using $\g$ itself. But this is precisely not the physical metric in the laboratory situation for electromagnetic analogue space-times. (Were one to prefer electromagnetism in the formalism of differential forms, the issue would be hidden in the Hodge star operator and its dependence on a metric, $\star_g \neq \star_{\g}$, as it depends both on the (inverse) metric and its determinant.) Thus, naturally, the laboratory will not observe the vacuum electrodynamics of $(U,\g)$. To emphasize this, compare equation~\eqref{eq:wave2} with the following vacuum wave equation of the physical background/laboratory metric $g$ itself:
	\begin{equation}
	\ed{\sqrt{\det g}}\partial_a \kl{\sqrt{\det g}\; g^{ad} g^{cb} \kl{\partial_d A_c - \partial_c A_d}}= J^b_\text{eff}.
	\end{equation}
	In order to interpret the electrodynamics in the medium in terms of the background metric's Levi-Civita connection, it is worthwhile to remind ourselves of the non-metricity tensor
	\begin{equation}
	q_{bca} \defi \nabla_a^\text{lab} g^\text{eff}_{bc}.\label{eq:defnonmetricity}
	\end{equation}
	While it makes the notation denser, we will from now on include the index \enquote{lab} to simplify the interpretation of the equations to come. 
	Let us denote with $\genfrac\{\}{0pt}{2}{a}{bc}_{\text{eff}}$ the Levi-Civita connection of the analogue metric, and with $\left[\Gamma^\text{lab}\right]^c_{ab}$ the Levi-Civita connection of the background space-time's metric $g$. Then we can write (see \cite{Schouten}):
	\begin{equation}
	\left[\Gamma^\text{lab}\right]^c_{ab} = \genfrac\{\}{0pt}{0}{c}{ab}_{\text{eff}} + \underbrace{\gi^{cm}\kl{\ed{2}\kle{q_{abm}-2q_{m(ab)}}}}_{\ifed \Delta^c{}_{ab}}.\label{eq:GammaLV}
	\end{equation}
	As both connections appearing are symmetric, no torsion terms can occur.
	
	If we now rephrase the inhomogeneous, macroscopic Maxwell equation of the lab in terms of the effective connection, we get instead of equation~\eqref{eq:wave} the following equation:
	\begin{align}
	\nabla_{a}^\text{lab} (Z^{abcd} F_{cd}^\text{lab}) = J^b &+\Delta^a{}_{ma} Z^{mbcd} F_{cd}^\text{lab} + \Delta^b{}_{ma} Z^{amcd} F_{cd}^\text{lab} + \Delta^c{}_{ma} Z^{abmd} F_{cd}^\text{lab} \nonumber\\& + \Delta^d{}_{ma} Z^{abcm} F_{cd}^\text{lab} - \Delta^m{}_{ca} Z^{abcd} F_{md}^\text{lab} -\Delta^m{}_{da} Z^{abcd} F_{cm}^\text{lab}.\label{eq:Ztangle}
	\end{align}
	The meaning of this equation is that in order to interpret any actual measurement of the laboratory in terms of vacuum electrodynamics of the analogue space-time is far from trivial. 	
	
	\subsection{Making Use of the Complication}
	Rather than just being troublesome, equation~\eqref{eq:Ztangle} also holds some small advantage: If one is interested in the analogue of a vacuum space-time with electromagnetic field \emph{not} equipped with a symmetric connection $\Gamma^\text{eff}$, this is just a minor change. Before doing so, let us be clear that an introduction of torsion (the consequence of an antisymmetric part in a connection) \emph{will} break gauge invariance, as shown in \cite{HehlObukTorsionHowToEdyn} --- or drop out. To avoid the latter, instead of using
	\begin{equation}
		F_{ab} \defi (\dif F)_{ab}
				= \partial_a A_b - \partial_b A_a
	\end{equation}
	as definition of the field strength tensor, we will use
	\begin{equation}
		F_{ab}	\defi \nabla_a A_b - \nabla_b A_a = (\dif F)_{ab} - 2 T^m{}_{ab} A_m,
	\end{equation}
	where $T^c{}_{ab}$ is the torsion tensor, and the decomposition of the laboratory Levi-Civita connection $\Gamma^\text{lab}$ (metric compatible to $g$) changes to
	\begin{equation}
	\left[\Gamma^\text{lab}\right]^c_{ab} = \left[\Gamma^\text{eff}\right]^c_{ab} + \underbrace{\gi^{cm}\kl{\ed{2}\kle{q_{abm}-2q_{m(ab)}} + \kle{2T_{(ab)m}-T_{mba}}}}_{\ifed \tilde \Delta^c{}_{ab}}.\label{eq:GammaLVTorsion}
	\end{equation}
	With this redefinition of $\tilde \Delta^c{}_{ab}$ it is now possible to find a very simple formal extension of equation~\eqref{eq:Ztangle}:
	\begin{align}
	\nabla_{a}^\text{lab} (Z^{abcd} F_{cd}^\text{lab}) = J^b 
	& + \text{terms dependent on $\Gamma_\text{eff}$ and its coupling to }F^\text{eff} \nonumber\\&
	+\tilde\Delta^a{}_{ma} Z^{mbcd} F_{cd}^\text{lab} + \tilde\Delta^b{}_{ma} Z^{amcd} F_{cd}^\text{lab} + \tilde\Delta^c{}_{ma} Z^{abmd} F_{cd}^\text{lab} 
	\nonumber\\ &
	+ \tilde\Delta^d{}_{ma} Z^{abcm} F_{cd}^\text{lab} - \tilde\Delta^m{}_{ca} Z^{abcd} F_{md}^\text{lab} -\tilde\Delta^m{}_{da} Z^{abcd} F_{cm}^\text{lab}.\label{eq:Ztangletorsion}
	\end{align}
	While this certainly is unlikely to make the actual interpretation of laboratory measurements any simpler, at least the formalism is not too different from the previous, Levi-Civita case. Nor would we be restricted to considering only Levi-Civita connections in equation~\eqref{eq:Ztangle}: Any symmetric, effective connection will do. Note that a non-symmetric connection would also entail an additional term $(\nabla^\text{eff}_a T^m{}_{cd})A^\text{eff}_m$ in the inhomogeneous Maxwell equation of the analogue space-time, as indicated in the first line of equation~\eqref{eq:Ztangletorsion}. However, it is again worth pointing out that torsion breaks gauge-invariance \cite{HehlObukTorsionHowToEdyn}. This introduces additional freedom for coupling the torsion terms to the Maxwell equations. For simplicity's sake we mirrored this freedom only in equation~\eqref{eq:Ztangletorsion}, and only in words. (We repeat: The coupling of electromagnetism to torsion also hinges on defining it through the antisymmetric, covariant derivative of $A$ instead of \enquote{just} the exterior derivative of $A$. The latter case precludes the coupling torsion in the first place, unless other measures for coupling torsion to electromagnetism are taken like moving to non-linear electrodynamics.)

	This results in a rather unexpected extension of the analogue space-time paradigm. While the experimental situation of this algebraic analogue certainly will not be simple, it holds the promise of being easily extended to more general theories of (an effective or analogue) gravity. It is possible to use the same formalism for a vast variety of different connections. The original title of the field, \enquote{analogue gravity} thus becomes more appropriate, as the additional, gravitational degrees of freedom captured in the connection, in non-metricity and torsion, could appear explicitly in the analogue. A thorough study of these analogues of modified theories of gravity will be future work. The only caveat on the preceding analysis is that the labelling discussed before has not been addressed at this point. It is mostly an exercise in the chain rule and notation, albeit an intricate one.
	
	\subsection{Example: The Hawking Effect}
	In order to better understand the difference between the coordinates of the analogue space-time and the corresponding labelling coordinates of the laboratory better, let us have a quick look at the Hawking effect. This will provide an explicit example for the cartographic distortions alluded to earlier and in \cite{TrafoOpticsCartographDistort}. The most general way to give the Hawking temperature is in terms of the surface gravity $\kappa$ at the exterior horizon (here located at $r=r_+$):
	\begin{equation}
	T = \frac{\hbar}{2\pi} \kappa(r_+).\label{eq:HawkingT}
	\end{equation}
	As one of the goals of the analogue space-time program is to make the Hawking effect experimentally accessible, the natural question in the present context is how to relate the Hawking temperature to the electromagnetic properties of the black hole space-time mimic. Put differently, one is looking for the functional relation
	\begin{equation}
	\kappa(r_+) = \kappa(\epsilon,\mu,\zeta).
	\end{equation}
	Identifying the surface gravity for an arbitrary effective metric on this level is equivalent to the question what the most general form of a time-like Killing vector field on \emph{some} effective metric is. It is doubtful that this can be done in closed form. Therefore, we will focus our attention on one particular type of metric, namely a general spherically symmetric and static one. Afterwards, we specialise to the Schwarzschild case contained in this type of metric. We assume the form of the metric to be
	\begin{equation}
	\dif s^2 = -f(r) \dif t^2 + f(r)^{-1} \dif r^2 + r^2\dif \Omega^2,\label{eq:metricwithf}
	\end{equation}
	where $f(r)$ is some function depending only on the radius and providing the metric in question with a (outer) horizon. For this metric, the time-like Killing vector field is easily found and given --- it is identical to $\partial_t$, and corresponds to translation in the coordinate time $t$. Thus we know that \cite{Poisson}
	\begin{subequations}
		\begin{align}
		\kappa(r_+) &= \sqrt{-\frac{t^{a;b}\; t_{a;b}}{2}},\\
		&=\ed{2} f'(r_+).\label{eq:kappaf}
		\end{align}
	\end{subequations}
	
	In reference \cite{CovEffMetrics}, we derived through an orthogonal decomposition expressions of the effective metric $\g$ in terms of the material's optical properties, $\epsilon, \mu, \zeta$. They read:
	\begin{subequations}\label{eq:g}
		\begin{align}
		\kl{g_\text{eff}}_{ab} &= -\sqrt{\frac{-\det(g^{\bullet\bullet})}{\pdet(\epsilon^{\bullet\bullet})}}\, V_a V_b + \sqrt{\frac{\pdet(\epsilon^{\bullet\bullet})}{-\det(g^{\bullet\bullet})}} [\epsilon^{\bullet\bullet}]^\#_{ab},\label{eq:geffeps0}\\
		&= -\sqrt{\frac{-\det(g^{\bullet\bullet})}{\pdet(\mu^{\bullet\bullet})}}\, V_a V_b + \sqrt{\frac{\pdet(\mu^{\bullet\bullet})}{-\det(g^{\bullet\bullet})}} \mu^{-1}_{ab}.\label{eq:geffmu0}
		\end{align}
	\end{subequations}
	Here $\#$ denotes the Moore--Penrose pseudo-inverse \cite{GenInverses}, and $\mu = \mi^\#$. The occurrence of the pseudo-inverse is related to the four-orthogonality of $\epsilon, \mi$ and $\zeta$ to $V$: As this means that $V$ is in the kernel of the corresponding linear maps, they cannot have a well defined matrix inverse. The Moore--Penrose pseudo-inverse circumvents this issue. Similarly, the pseudo-determinant $\pdet(A)$ is the product of all non-zero singular values (eigenvalues). (For references on the rarely used concept of pseudo-determinant we refer to our earlier paper \cite{CovEffMetrics}.) It can be seen as a notion of a determinant appropriate for the Moore--Penrose pseudo-inverse, but is an independent (though relatable) concept. The definition of the Moore--Penrose pseudo-inverse does not rely on the pseudo-determinant, nor does the pseudo-determinant require the introduction of the Moore--Penrose pseudo-inverse. 
	
	Equivalently to equation~\eqref{eq:g}, one could make use of a $3+1$ style decomposition:
	\begin{subequations}
		\begin{align}
		[g_\text{eff}]_{ab} &= 
		\begin{pmatrix}
		-\sqrt{\det(\mu^{\circ\circ})}^{-1} &  \mu^{-1}_{\,jk}\beta^k\\[.5em]
		\mu^{-1}_{\,ik}\beta^k 
		&\sqrt{\det(\mu^{\circ\circ})}\kl{\mu^{-1}_{\,ij} - ( \mu^{-1}_{\,ik} \beta^k) ( \mu^{-1}_{\,jl}\beta^l)}
		\end{pmatrix},\label{eq:g3+1mu}\\
		&= \begin{pmatrix}
		- \sqrt{\det \epsilon^{\circ\circ}}^{-1}(1-\epsilon_{kl}^{-1}\beta^k\beta^l) & \epsilon_{jk}^{-1}\beta^k \\[.5em]
		\epsilon^{-1}_{ik}\beta^k & \sqrt{\det\epsilon^{\circ\circ}}(\epsilon^{-1}_{ij})
		\end{pmatrix},\label{eq:g3+1eps}\\
		\text{with~} \beta^m &\mathrel{\mathop:}= \sqrt{\det(\epsilon^{\circ\circ})} \; \eps^{mk}{}_i \; \epsilon^{-1}_{\,jk}\; \zeta^{ij}.\nonumber
		\end{align}
	\end{subequations}
	
	Combining the metric~\eqref{eq:metricwithf} with the results of equation~\eqref{eq:geffeps0} or equation~\eqref{eq:geffmu0} then gives that
	\begin{subequations}\label{eq:f}
		\begin{align}
		f(r) & = \sqrt{\det g\; \det \mu^{-1}},\\
		&=\sqrt{\det g \; \det \epsilon^{-1}},
		\end{align}
	\end{subequations}
	We could have also started from either equation~\eqref{eq:g3+1mu} or equation~\eqref{eq:g3+1eps}. But we have to be very careful: In deriving either, heavy use of the conformal freedom in $3+1$-dimensional electrodynamics was made. This would utterly obfuscate the inclusion of factors of determinants of the background metric $g$.
	
	Then inserting this result~\eqref{eq:f} in equation~\eqref{eq:kappaf} allows us to identify the Hawking temperature in terms of either permeability or permittivity:
	\begin{subequations}\label{eq:HawkingEff}
		\begin{align}
		T_\text{H} &= \left.\kl{\frac{\partial}{\partial r_\text{eff}}\frac{\hbar}{4\pi \sqrt{\det g\;\det\epsilon}}}\right\lvert_{r_\text{eff}=r_+},\\
		&= \left.\kl{\frac{\partial}{\partial r_\text{eff}}\frac{\hbar}{4\pi \sqrt{\det g \;\det\mu}}}\right\lvert_{r_\text{eff}=r_+}.
		\end{align}
	\end{subequations}
	It is worth pointing out that the constitutive tensors are invariant under conformal transformations of the effective metric $g_\text{eff}$. Naturally, this means that also the determinant of $\epsilon$ and $\mu$ are conformally invariant, and hence the Hawking temperature as given in equations~\eqref{eq:HawkingEff}. This is in full agreement with the result of Jacobson and Kang, \cite{JacobsonKang93}, that the Hawking temperature in general should be conformally invariant. In preparation for the next step, we already started distinguishing between laboratory and effective coordinates.
	
	Let us be more concrete about the effective metric under consideration, and choose $f(r)=1-2M/r$ in equation~\eqref{eq:metricwithf}, as appropriate for an effective Schwarzschild black hole (but setting $G_\text{N}=1$).\footnote{As the occurrence of natural constants in equation~\eqref{eq:HawkingT} cancels all length dimensions to leave only Kelvin, we will not have to worry about scaling of natural constants in what follows.} The \enquote{mass} $M$ is best understood as a mere parameter of the effective space-time being mimicked in the laboratory. It will not correspond to any actual physical mass. The constitutive tensors of the corresponding meta-material mimic are then easily calculated. They are:
	\begin{subequations}
		\begin{align}
		\epsilon^{ab} &= \begin{pmatrix}
		1 & 0 & 0\\ 0 & \ed{r(r-2M)} & 0 \\ 0 & 0 & \ed{r(r-2M)\sin^2\theta}
		\end{pmatrix},\\
		\mu^{ab} &= \begin{pmatrix}
		1 & 0 & 0\\ 0 & \ed{r(r-2M)} & 0 \\ 0 & 0 & \ed{r(r-2M)\sin^2\theta}
		\end{pmatrix},\\
		\zeta^a{}_b &= 0.
		\end{align}
	\end{subequations}
	However, the temperature as given in equations~\eqref{eq:HawkingEff} is unlikely to be immediately measured. In a laboratory, as described in section~\ref{sec:coords}, the coordinates would be different to those of the effective space-time. In particular, the relation is unlikely to be a four-dimensional conformal transformation. To illustrate this, let us suppose we stretch or shrink the radial coordinate when going from laboratory coordinates to those of the effective space-time,
	\begin{subequations}
		\begin{align}
		(t,r,\theta,\phi)_\text{lab} \qquad&\longrightarrow \qquad (t,a\,r_\text{lab},\theta,\phi)_\text{eff},\\
		r_\text{eff} &= a \,r_\text{lab}.
		\end{align}
	\end{subequations}
	The example of the Schwarzschild metric would now read in the stretched coordinates
	\begin{equation}
	\dif s^2_\text{eff} = - \kl{1-\frac{2M}{a\,r_\text{lab}}}\dif t^2 + a^2 \kl{1-\frac{2M}{a\,r_\text{lab}}}^{-1} \dif r_\text{lab}^2 + a^2\, r_\text{lab}^2\dif \Omega^2.
	\end{equation}
	There are several important points to be made here: First note, that while it \emph{could} be viewed as a simple coordinate transformation, this would ignore the physical significance of $r_\text{lab}$. This is, after all, the distance measured in a laboratory, thus lending particular significance to it. Second, note that this does \emph{not} correspond to a conformal transformation of an effective Schwarzschild metric. Third, note that when performing this calculation, it is absolutely vital to not fix conformal factors to simplify calculations, as the scale $a$ will break this choice. Since only the $r$-coordinate differs in the laboratory coordinate system from the corresponding one in the effective space-time, we only place appropriate index labels (\enquote{eff}) on $r$. This then corresponds to electromagnetic tensors
	\begin{subequations}
		\begin{align}
		\epsilon^{ab} &= \begin{pmatrix}
		a & 0 & 0 \\ 0 & \frac{a^2}{(a\,r_\text{lab}-2M)r_\text{lab}}& 0 \\ 0&0&\frac{a^2}{(a\,r_\text{lab}-2M)r_\text{lab}\sin^2\theta}
		\end{pmatrix},\\
		\mu^{ab} &= \begin{pmatrix}
		a & 0 & 0 \\ 0 & \frac{a^2}{(a\,r_\text{lab}-2M)r_\text{lab}}& 0 \\ 0&0&\frac{a^2}{(a\,r_\text{lab}-2M)r_\text{lab}\sin^2\theta}
		\end{pmatrix},\\
		\zeta^a{}_{b} &= 0.
		\end{align}
	\end{subequations}
	
	If we now follow through with the calculation of the Hawking temperature, we see that for $a=1$, which is the \enquote{standard} Schwarzschild metric, equations~\eqref{eq:HawkingEff} correctly evaluate to
	\begin{equation}
	T_\text{H} = \frac{\hbar}{8\pi M}.
	\end{equation}
	However, if we keep $a$ arbitrary, we now get (after a bit of calculating)
	\begin{equation}
	T_\text{H} = \frac{\hbar}{8\pi a^{9/2} M}.
	\end{equation}
	This is the temperature measured in the laboratory. We give some values for the temperature as they would be measured in the laboratory in table~\ref{tab:TH}, setting the radius of the analogue event horizon to be $\SI{10}{\centi\meter}$ for a range of different values of $M$. As we see, if we attempt to fit a microscopic black hole geometry inside a tabletop laboratory setting, the scale factor will actually increase the Hawking temperature (as $a\ll 1$) to ludicrous values. On the other hand, if we attempted to mimic an actual astrophysical black hole (at mass regimes we know or assume their existence), we would have to blow the radial coordinate up with $a\gg 1$, greatly diminishing the Hawking temperature. Already the temperature of a solar mass black hole becomes laughably small, not to mention that of a black hole of the mass of Sagitarius A$^*$, here taken to be $M_{\text{Sgr A}^*}=\num{4e6} M_\odot$: A temperature in the range of $\SI{100}{\pico\kelvin}$ can be achieved in the context of Bose--Einstein condensates,\footnote{For this, we refer to Knuuttila's PhD thesis \cite{LowestTemp}. The upcoming results of the Cold Atom Laboratory aboard the International Space Station will likely further lower this particular record, see \url{https://coldatomlab.jpl.nasa.gov/}, accessed at 00:08am, March $8^\text{th}$ 2019.} but only much more modest $\SI{6}{\milli\kelvin}$ are achievable for macroscopic objects,\footnote{Experimentally achieved by the CUORE collaboration at the INFN Gran Sasso National Laboratory, see \url{https://www.interactions.org/node/12905}, accessed at 00:09am, March $8^\text{th}$ 2019.} while the lowest naturally occurring temperatures observed to date is about $\SI{1}{\kelvin}$ in the Boomerang nebula\footnote{Source: \url{https://apod.nasa.gov/apod/ap071228.html}, accessed at 00:09am, March $8^\text{th}$ 2019.} due to gas expansion. 
	
	However, notice that one still wants to keep to astronomical objects as effective masses --- already the effective temperature of a black hole of Earth mass is reasonable, and even the temperature range of a black hole with Venus' mass might be achievable. Depending on the experimental set-up (cooling, accuracy of temperature measurements) even the comparatively small values for Uranus might still be realisable. We included the proton mass and $\SI{1}{\kilogram}$ only for illustrative purposes. Modelling an elementary particle by black hole space-times would be riddled with \emph{many} more issues in the first place --- including naked singularities, and that it is \emph{charged} and \emph{has spin} while we are calculating values for uncharged, non-rotating Schwarzschild black holes. (This list can easily be extended, including but not limited to: Much less than a bit of information in the Bekenstein entropy, energy peak frequencies higher than the rest mass, all the problems galore in league with naked singularities, not to mention that its (Kerr--Newman) ring singularity would have a radius already excluded by experiments on electrons\dots)
	
	Let us reiterate: While \enquote{masses} slightly above the mass of the Earth give experimentally achievable temperatures, \emph{these are the temperatures of the analogues}. A real, \emph{astrophysical} object of comparable mass would have the well-known, low Hawking temperatures given in the third column of table~\ref{tab:TH}. Likewise, the experimental situation would not be concerned with any masses as given in the first column. These \enquote{masses} only have meaning in the analogue space-time (as we work with an effective Schwarzschild geometry, it does not matter which mass concept precisely we invoke, so \enquote{effective} ADM masses are sufficient). The only physical mass involved will be that of the optical medium used as an effective space-time.
	
	Lastly, and probably most importantly, the calculation shown here can only be considered heuristic in nature: In order to validate these results, one would have to perform a quantum field theoretic analysis based on the field modes of the analogue, \emph{i.e.}, the electromagnetic field of the laboratory's material mimicking the space-time.
	
	\begin{table}
		\flushright
		\begin{tabular}{||c|c|c|c||}\hline\hline
			$M[\si{\kilo\gram}]$ & $r_\text{H}[\si{\meter}]$ & $T_\text{H}[\si{\kelvin}]$  & $T_\text{lab}[\si{\kelvin}]$\\\hline
			$M_p = \num{1.6726e-27}$ & $\num{2.4841e-54}$ & $\num{7.3355e49}$ & $\num{3.8651e286}$\\
			$\num{1}$ & $\num{1.4852e-27}$ & $\num{1.2269e23}$ & $\num{2.0693e139}$\\
			$M_{\leftmoon} = \num{7.342e22}$ & $\num{1.0904e-4}$ & $\num{1.6711}$ & $\num{3.5796e13}$\\
			$M_{\mars} = \num{6.4171e23}$& $\num{9.5306e-4}$ & $\num{0.1912}$ & $\num{2.3738e8}$\\
			$M_{\venus} = \num{4.8685e24}$ & $\num{7.2306e-3}$ & $\num{2.5202e-2}$ & $\num{3428.8}$\\
			$M_\oplus = \num{5.9736e24}$ & $\num{8.8719e-3}$ & $\num{2.0539e-2}$ & $\num{1113.1}$\\
			$M_{\uranus} = \num{8.6832e25}$ & $\num{0.1290}$ & $\num{1.4130e-3}$ & $\num{4.4986e-4}$\\
			$M_\odot = \num{1.9886e30}$ & $\num{2953.4}$ & $\num{6.1700e-8}$ & $\num{4.7191e-28}$\\
			$M_{\text{Sgr A}^*} = \num{7.9542e36}$ & $\num{1.1813e10}$ & $\num{1.5425e-14}$ & $\num{2.3043e-64}$\\\hline\hline
		\end{tabular}
		\caption{Table comparing the Hawking temperature $T_\text{H}$ as \enquote{observed} within the effective (Schwarzschild) space-time itself with the actually observed temperature $T_\text{lab}$ for different values of the black hole mass $M$. The value of $a$ is chosen such that the radius of the event horizon in the laboratory is $\SI{10}{\centi\meter}$. Thus, $a$ is simply $\num{10}$ times the value of $r_\text{H}$ in meters.}
		\label{tab:TH}
	\end{table}
	\FloatBarrier
	\section{An Analytic Analogue}
	In this section, we shall give an alternative access to analogue space-time models building on electromagnetism. Unlike in the previous section, this time we will work in what we call an \emph{analytic} analogy. Presently, our analytic analogy operates by linking the dynamics of partial differential equations (PDEs) in GR with those of the PDEs in electromagnetism. It is thus an instance of the vast majority of analogue space-time models, most being of this \emph{analytic} kind. It can be argued that Gordon's original motivation for what we now called a model for an \emph{algebraic} analogue space-time was to create such an analytic analogue. However, the preceding sections should have convinced the reader that only the raising and lowering of indices with an effective metric on the level of the Lagrangian was mimicked, not the full PDE of electromagnetism in an effective vacuum space-time. 
	
	Our new, \emph{analytic} analogy will lead to refractive index profiles analogous to aspects of wave propagation in a curved space-time. More explicitly, we will derive a stratified refractive index profile for which a mode of frequency $\omega$ will propagate exactly in the same way as a radially propagating mode of the same frequency would in the curved space-time. While this analytic \enquote{analogy} is only mimicking one aspect of the full curved space-time physics (unlike what a full \enquote{analogue} hopes to achieve), the resulting index profiles might be more readily available than the complicated, tensorial electromagnetic properties $\epsilon, \mu, \zeta$.
	
	This approach was inspired by the work of Heading and Westcott \cite{HeadingHyperGeo,WestcottN,HeadingSingFreeIndices}. Their idea is to find exactly solvable refractive index profiles (using available special functions) in horizontally \cite{HeadingHyperGeo} or spherically \cite{WestcottN} stratified continuous media by looking at the wave equation for the part of the electric/magnetic field propagating orthogonally to the stratification layers. Much of their analysis was dedicated to finding further, exactly solvable profiles beyond those known at their time. In our case, the task is framed in a slightly different way: Starting from the wave equation for massless perturbations of the Kerr space-time, we use its separability to reduce the problem to a system of ordinary differential equations. Looking at the \enquote{radial} equation, we then try to recognize the form of a one-dimensional Helmholtz equation for stratified media, which can be written in terms of a refractive index depending on one variable.
	
	There does exist an extension of the methods of Heading and Westcott in \cite{HeunRefIndex} to tackle second order Fuchsian equations with four singular points, that is, to functions of the Heun class \cite{KristenssonODE,SlavyanovLay00,RonveauxHeun}. We shall follow a slightly different approach which will be less general, and more immediately related to our intended application in the context of analogue space-times.
	
	Following \cite{Bremmer49}, and \cite{MoonSpencer}, such \enquote{target} Helmholtz equations can be found, for example, either in spherical symmetry, or in cylindrical symmetry. We shall focus on these. Regarding the horizontal stratification (\emph{i.e.}, in $z$-direction) in Cartesian coordinates, the classic text by Born and Wolf \cite{BornWolf} gives the resulting Helmholtz equation on pages~55--56 in terms of scalar permeability $\mu$ and refractive index $n$. If the stratification of the refractive index $n$ is pushed solely into the permittivity this is turned into a special case of the discussion to follow below. Once the required formalism is in place, we will give the final detail regarding this.
	
	For spherical coordinates the scalar Helmholtz equation can be derived as a scalar equation to be fulfilled by a radially directed Hertzian vector $\vec{\Pi}$. This gives two possible scalars $\Pi$ --- one for an electrical dipole in radial direction, $\Pi_{\text{e}}$, and one for the case of a magnetic dipole in radial direction, $\Pi_{\text{m}}$. More concretely, the corresponding electromagnetic fields are determined by
	\begin{subequations}
		\begin{align}
		\vec{E} &= \frac{c}{\omega^2 n^2(r)} \nabla \times \nabla \times (\vec{r} \omega/c \Pi_{\text{e}}) e^{-i\omega t},\\
		\vec{H} &= -i \nabla \times (\vec{r} \omega/c \Pi_{\text{e}}) e^{-i\omega t}
		\end{align}
	\end{subequations}
	for the electric case, and for the magnetic one in the following way:
	\begin{subequations}
		\begin{align}
		\vec{E} &= i\frac{c}{\omega n^2(r)} \nabla \times (\vec{r} (\omega/c)^2 \Pi_{\text{m}}) e^{-i\omega t},\\
		\vec{H} &= c^2 \frac{c}{\omega^2 n^2(r)} \nabla \times \frac{\nabla \times (\vec{r} \omega/c \Pi_{\text{e}})}{\omega^2 n^2(r)} e^{-i\omega t}.
		\end{align}
	\end{subequations}
	The spherical scalar Helmholtz equation with refractive index $n(r)$ then reads
	\begin{equation}\label{eq:scHeqnEsph}
	\frac{\dif^2}{\dif r^2}\Pi_{\text{e}} + \kl{\frac{\omega^2n^2(r)}{c^2} - n(r)\frac{\dif^2}{\dif r^2}\kl{\ed{n(r)}} - \frac{l(l+1)}{r^2}}\Pi_{\text{e}}=0
	\end{equation}
	for the electric case, and
	\begin{equation}\label{eq:scHeqnMsph}
	\frac{\dif^2}{\dif r^2}\Pi_{\text{m}} + \kl{\frac{\omega^2n^2(r)}{c^2} - \frac{l(l+1)}{r^2}}\Pi_{\text{m}}=0
	\end{equation}
	for the magnetic one. Note the occurrence of the separation constant $l(l+1)$ from the separation of variables of the scalar Helmholtz equation. These $l$ can be given the usual interpretation in terms of angular momentum quantum numbers.
	
	In the case of cylindrical coordinates $(r,\theta,z)$, we take a slightly different path and we start by looking at fields independent of $z$ (which is justified as it is the $r$-direction where we are looking for stratification): The $z$-components for $\vec{H}$ and $\vec{E}$ then can be decomposed as
	\begin{subequations}
		\begin{align}
		E_z &= \sum_{l=0}^{\infty} a_l r^{-1/2} f_l(r) \cos(l\theta),\\
		H_z &= \sum_{l=0}^{\infty} b_l r^{-1/2} g_l(r) \cos(l\theta).
		\end{align}
	\end{subequations}
	Here, $a_l, b_l$ are constants, and $f_l(r)$, $g_l(r)$ fulfil the following scalar Helmholtz equations \cite{BurmanCylProp,WestcottCylProp}:
	\begin{alignat}{2}
	\frac{\dif^2 f_l(r)}{\dif r^2} &+ \kl{\frac{\omega^2}{c^2}n^2(r) - \frac{l^2-\ed{4}}{r^2}}f_l(r) &&= 0,\label{eq:scHeqnEcyl}\\
	\frac{\dif^2 g_l(r)}{\dif r^2} &+ \kl{\frac{\omega^2}{c^2}n^2(r) -\frac{\omega}{cr}\frac{\dif}{\dif r}\kl{r\frac{\dif}{\dif r}\kl{\frac{c}{\omega n(r)}}} - \frac{l^2-\ed{4}}{r^2}}g_l(r) &&= 0.\label{eq:scHeqnMcyl}
	\end{alignat}
	
	In principle, one could also try to look at the scalar Helmholtz equation in spheroidal symmetry --- a separation of variables is also possible in this case. However, special care has to be taken when considering this last case: As the \emph{vector} Helmholtz equation is not separable in spheroidal coordinates, the origin of a spheroidal \emph{scalar} Helmholtz equation in this essentially electromagnetic analogy would have to be carefully justified. It is not unlikely, though, that in a different physical setting an application also for the spheroidal scalar Helmholtz equation can be found along the lines of the following discussion for the first two cases.
	
	We will be particularly interested in the cases of equation~\eqref{eq:scHeqnMsph}, and equation~\eqref{eq:scHeqnEcyl}. Both are of the form
	\begin{equation}
	\frac{\dif^2 f(r)}{\dif r^2} + \kl{\frac{\omega^2}{c^2}n^2(r) - \frac{D}{r^2}}f(r) = 0,\label{eq:refindD}
	\end{equation}
	though for different, but constant values of $D$ and different dependent functions $f(r)$. The horizontal stratification à la Born and Wolf, \cite{BornWolf}, with $\mu$ chosen $z$-independent, appears here as a special case of setting $D$ equal to $0$. This case is covered by the spherical Helmholtz equation for $M$, when setting $l=0$. (From a more technical point of view, one could also change the variable in \cite{BornWolf}, equation~(6), page~56, to get rid of the first-derivative piece with standard methods from the theory of second order ODEs; but we are neither interested in keeping separately track of $\epsilon$ and $\mu$, nor would the resulting ODE have quite the appearance of equation~\eqref{eq:refindD} on which we shall focus. If we kept both $\epsilon(r)$ and $\mu(r)$ yet another new method for creating an analogy would open up.)
	
	The other two Helmholtz equations~(\ref{eq:scHeqnEsph},\ref{eq:scHeqnMcyl}) will not be further considered: While our method for finding an analogy to (a part of) the wave equation of the Kerr space-time would still work, it will not be as convenient. Equations~(\ref{eq:scHeqnMsph},\ref{eq:scHeqnEcyl}) will result in a refractive index as some explicit expression involving a square root. Equations~(\ref{eq:scHeqnEsph},\ref{eq:scHeqnMcyl}), in contrast, would result in a needlessly complicated first order ODE $n(r)$ would have to fulfil.
	
	To turn any of these scalar Helmholtz equations into a simple space-time analogy, we have to take a closer look at the wave equation for massless fields in the Kerr space-time. This wave equation is given by the Teukolsky equation, and we will be following its presentation in reference \cite{FroNo98},
	\begin{align}\label{eq:Teukolsky}
	0=& \kle{\frac{(r^2+a^2)^2}{\Delta} - a^2\sin^2\theta} \frac{\partial^2\Psi}{\partial t^2} + \frac{4Mar}{\Delta}\frac{\partial^2 \Psi}{\partial t \partial \phi} + \kle{\frac{a^2}{\Delta} - \ed{\sin^2\theta}}\frac{\partial^2\Psi}{\partial \phi^2}\nonumber\\
	&-\Delta^{-s} \frac{\partial}{\partial r}\kl{\Delta^{s+1}\frac{\partial\Psi}{\partial r}} - \ed{\sin\theta} \frac{\partial}{\partial\theta} \kl{\sin\theta \frac{\partial \Psi}{\partial \theta}} - 2s\kle{\frac{a(r-M)}{\Delta} + i\frac{\cos\theta}{\sin^2\theta}}\frac{\partial\Psi}{\partial\phi}\nonumber\\
	&-2s \kle{\frac{M(r^2-a^2)}{\Delta} - r -ia\cos\theta}\frac{\partial\Psi}{\partial t} + (s^2\cot^2\theta - s)\Psi.
	\end{align}
	Here, $s$ is the so-called spin-weight, while the other terms are defined as in the line element of the Kerr solution:
	\begin{subequations}
		\begin{equation}\label{eq:Kerr}
		\dif s^2 = - \frac{\Delta}{\rho^2} \kl{\dif t - a \sin^2 \theta \dif \phi}^2 + \frac{\sin^2\theta}{\rho^2}\kl{(r^2+a^2)\dif \phi - a \dif t}^2 + \frac{\rho^2}{\Delta} \dif r^2 + \rho^2\dif \theta^2,
		\end{equation}
		where
		\begin{gather}
		\Delta \defi r^2 - 2Mr +a^2,\\
		\rho^2 \defi r^2 + a^2 \cos^2 \theta,\\
		a \defi L/M,
		\end{gather}
	\end{subequations}
	and $M$ is the mass of the black hole, while $L$ is its angular momentum.
	
	The Teukolsky equation~\eqref{eq:Teukolsky} describes massless (classical) fields of helicity $s$. Usually, these are interpreted as perturbations of the corresponding metric, and play a role in stability analyses of the metric as a solution of the Einstein equations. We, however, shall in this section consider them as an approximation to the propagation of (massless) radiation emitted or scattered by the black hole --- this is just a rephrasing of the question that generated the Teukolsky equation in the context of black hole stability analysis. Correspondingly, it is just as important for scattering problems, including, but not limited to, the finding of quasi-normal modes.
	
	In order to further distinguish the analogy from the physical laboratory in which the Helmholtz equation occurs, we omit physical constants from the former, while keeping the speed of light $c$ in the latter.
	
	While more general versions of the Teukolsky equation exist --- the separation of variables works in any Petrov type $D$ space-time (see \cite{GalErt89,HeunTypeD}; Kodama and Ishibashi further generalised this to higher dimensions in work culminating in \cite{KodamaIshibashiMasterEqn3}.) --- we shall be concerned as a proof of concept with the special case of the Kerr solution, as described above. These more general differential equations can be distinguished from the version above by referring to them as the \enquote{(Teukolsky) master equation} instead.
	
	The Teukolsky equation~\eqref{eq:Teukolsky} can be separated using the following mode decomposition of $\Psi$:
	\begin{subequations}
		\begin{align}
		\Psi &= \sum_{\ell =0}^{\infty} \sum_{m=-\ell}^{\ell} \vphantom{\Psi}_s\Psi_{\ell m},\\
		&= \sum_{\ell =0}^{\infty} \sum_{m=-\ell}^{\ell} \vphantom{R}_sR_{\ell m}(r,\omega) \vphantom{Z}_sZ_{\ell m}(\theta,\phi) e^{-i\omega t},\\
		&= \sum_{\ell =0}^{\infty} \sum_{m=-\ell}^{\ell} \vphantom{R}_sR_{\ell m}(r,\omega) \ed{\sqrt{2\pi}}\vphantom{S}_sS_{\ell m}(\theta) e^{im\phi} e^{-i\omega t}.
		\end{align}
	\end{subequations}
	
	Inserting this ansatz, we get the following equation for $\vphantom{S}_sS_{\ell m}(\theta)$:
	\begin{align}
	0=&\ed{\sin\theta} \frac{\dif}{\dif \theta}\kl{\sin\theta\frac{\dif \vphantom{S}_sS_{\ell m}(\theta)}{\dif \theta}} + \kl{E_{\ell m} -s^2}\vphantom{S}_sS_{\ell m}(\theta)\nonumber\\
	& + \kl{a^2\omega^2\cos^2\theta - \frac{m^2+2ms\cos\theta + s^2\sin^2\theta}{\sin^2\theta} - 2a\omega s \cos\theta }\vphantom{S}_sS_{\ell m}(\theta), 
	\end{align}
	while for $\vphantom{R}_sR_{\ell m}(r)$ we have
	\begin{align}\label{eq:radialTeukolsky}
	0=& \Delta^{-s} \frac{\dif}{\dif r} \kl{\Delta^{s+1} \frac{\dif \vphantom{R}_sR_{\ell m}(r)}{\dif r}} + \kl{s(s+1) - E_{\ell m} - a^2\omega^2 + 2am\omega}\vphantom{R}_sR_{\ell m}(r)\nonumber\\
	& + \kl{\frac{[(r^2+a^2)\omega - am]^2 - 2is(r-M)[(r^2+a^2)\omega-am]}{\Delta} + 4is\omega r }\vphantom{R}_sR_{\ell m}(r),
	\end{align}
	where $E_{\ell m} = E_{\ell m}(a\omega)$ is the separation constant. The functions $\vphantom{S}_sS_{\ell m}(r)$ are known as spin-weighted spheroidal harmonics, which become spherical harmonics in the case of the Schwarzschild solution where $a=0$ \cite{EVandEFspheroidal}.
	
	The values of $E_{\ell m}$ will be taken from \cite{TeukolskyPress74}, where a table of polynomial approximations in $a\omega$ can be found on page~454. Specifically, we will test two modes for photons. The choice fell on photons as they will still retain some fundamental relevance in this particular analogy --- after all, it is built on macroscopic electromagnetism. The modes we want to look at are $\ell=1, m=0$ and $\ell=6,m=3$. This choice is arbitrary, any other mode would work equally well. The corresponding values for the separation constant $E_{\ell m}$ are
	\begin{alignat}{5}
	E_{1,0} &= 2 &&+ 0.00281 a \omega &&- 0.413370 a^2 \omega^2 &&+ 0.021476 a^3 \omega^3 &&- 0.0335098 a^4 \omega^4 \nonumber\\&&&+ 0.0025402 a^5 \omega^5 &&+ 0.00032399 a^6 \omega^6,&&&&\\
	E_{6,3} &= 42&&- 0.14285 a \omega &&- 0.385851 a^2 \omega^2 &&+ 0.003204 a^3 \omega^3 &&+ 0.0002062 a^4 \omega^4 \nonumber\\&&&+ 0.0000197 a^5 \omega^5 &&+ 0.00000229 a^6 \omega^6.&&&&
	\end{alignat}
	The range of these polynomial fits is $0\leq a\omega \lesssim 3$.
	
	It is worth pointing out at this stage that we are \emph{not} looking at anything related to quasi-normal modes (QNM)~\cite{QNMLRR}. The questions resulting in the search for QNM are independent of our current aim. As a two-point connection problem in the sense of special function theory underlies QNM analyses (and thus more restrictive boundary conditions), the frequencies of QNM tend to be discrete. What we are looking at is rather the question of arbitrary incoming waves of arbitrary frequency. This corresponds to plane waves in the standard flat-space context. In this analogy, the conditions where QNM arise have the correspondence of the addition of boundary conditions to flat-space examples --- which also there can force eigenfrequencies to be discrete.
	
	Having separated the Teukolsky equation, we can now ask a more modest question in the same vein as our previous discussion of space-time analogues: Instead of looking at an analogue for the full wave propagation on a curved background, we restrict ourselves to a mode-by-mode analysis. Then the separation allows us to look for a one-dimensional analogy of a particular radial mode; each mode characterised by corresponding, specific separation constants. The analogy would then be simply given as a one-dimensional refractive index profile. However, and to our chagrin, the radial Teukolsky equation contains a \emph{complex} potential. While it is certainly not impossible to follow down the lines of complex refractive indices (and at least both Heading and Westcott later did, see \cite{HeadingComplex,WestcottComplex}) --- this would be a rather jarring contrast to the algebraic analogue space-time picture considered before. It bears repeating that if one allows the optical properties to be complex-valued, then the resulting dispersion would quickly destroy the Lorentzian analogy in the sense of the algebraic analogy. We ignore for a moment that wave equations in black hole space-times indeed exhibit absorption (due to backscattering), and thus would make complex refractive indices in the current, \emph{analytic} context not too surprising. (Again, this is different from the purely algebraic analogy of the previous section~\ref{sec:alg}!) To our knowledge, however, refractive indices are easier to manufacture to order if only the real part has to be matched. With the above complex potential, it therefore seems that we reached an impasse, or at least an unpleasant complication. 
	
	Luckily, however, at least for electromagnetic fields --- that is, $s=\pm1$ --- the Teukolsky equation allows for transformations turning the potential real, and thus available to our desired, one-dimensional space-time analogy~\cite{DetweilerPotential}. Hence, let us repeat this process here (while skipping some intermediate steps; for example, we shall only implicitly use the intermediate step of relating $\vphantom{R}_sR_{\ell m}(r)$ and $\vphantom{R}_{-s}R_{\ell m}(r)$ through the Teukolsky--Starobinsky identities to explain the origin of the following transformations):
	
	Instead of $\vphantom{R}_sR_{\ell m}(r)$, consider the following function:
	\begin{subequations}
		\begin{align}
		\vphantom{\chi}_1\chi_{\ell m}(r) \defi& p(r) \vphantom{R}_1R_{\ell m}(r) + \frac{p(r)\Delta}{\sqrt{4\lambda_{-1}^2 - 16a^2\omega^2 + 16a\omega m}} \vphantom{R}_{-1}R_{\ell m}(r),\\
		=&p(r)\kl{1 + \frac{\Delta}{\sqrt{4\lambda_{-1}^2 - 16a^2\omega^2 + 16a\omega m}}A(r)} \vphantom{R}_1R_{\ell m}(r)\nonumber\\
		&+\frac{p(r)\Delta}{\sqrt{4\lambda_{-1}^2 - 16a^2\omega^2 + 16a\omega m}}B(r)\frac{\dif\vphantom{R}_1R_{\ell m}(r)}{\dif r},
		\end{align}
		where
		\begin{align}
		p(r) \defi& \sqrt{\frac{\lambda_{-1}^2 - 4a^2\omega^2 + 4a\omega m}{\frac{2[am-\omega(r^2+a^2)]^2}{\Delta} -\lambda_{-1} + \sqrt{\lambda_{-1}^2 - 4a^2\omega^2 + 4a\omega m}}},\\
		\lambda_{-1} \defi& E + a^2\omega^2 -2a\omega m,\\
		A(r) \defi& \frac{2}{\Delta^2}\kle{2(am-\omega(r^2+a^2))^2 - \Delta(i\omega r + \lambda_{-1})},\\
		B(r) \defi& -4\frac{am-\omega(r^2+a^2)}{\Delta}i.
		\end{align}
	\end{subequations}
	With this redefined dependent variable it is now possible to rewrite the ODE for $r$ as
	\begin{equation}\label{eq:chiODE1}
	\frac{\dif^2 \vphantom{\chi}_1\chi_{\ell m}}{\dif r^2} + V(r,\omega)\, \vphantom{\chi}_1\chi_{\ell m} = 0,
	\end{equation}
	with the potential defined as
	\begin{equation}
	V(r,\omega) \defi \frac{[am-\omega(r^2+a^2)]^2}{\Delta^2} - \lambda_{-1} - \frac{[am-\omega(r^2+a^2)]p'' - 2r\omega p'}{(am-\omega[r^2+a^2])p}.
	\end{equation}
	
	At this stage it is already possible to make several comments: First, we notice that the complex variables have been pushed from the original ODE into the dependent variable $\vphantom{\chi}_1\chi_{\ell m}$. After this has been done, the complex nature of $\vphantom{\chi}_1\chi_{\ell m}$ can be ignored, as we are now looking only for solutions of equation~\eqref{eq:chiODE1}. If these solutions are complex is irrelevant: As the linear(!) differential operator has only real coefficients any non-trivially complex solution (meaning: both real and imaginary part are unequal 0) will necessarily lead to a second, linear independent solution formed from its complex conjugate --- and thus two linearly independent, \emph{real} solutions become available. Second, unlike the prototype equations~(\ref{eq:scHeqnEsph}, \ref{eq:scHeqnMsph}, \ref{eq:scHeqnEcyl}, \ref{eq:scHeqnMcyl}) the refractive index is now frequency-dependent. While the resulting dispersion is likely to complicate the experimental realisation, the absence of mode mixing (thanks to the separation) will alleviate this deficit somewhat.
	
	As an aside it is prudent to bear in mind that our analysis is founded on a single mode analysis. As such it will be difficult to check the results reported in \cite{SuperradianceOrNotBig,SuperradianceOrNot} which claim reflection of wave packets instead of super-radiance.
	
	The next step is to separate in equation~\eqref{eq:chiODE1} the terms corresponding to the refractive index in the target equations~(\ref{eq:scHeqnEsph}, \ref{eq:scHeqnMsph}, \ref{eq:scHeqnEcyl}, \ref{eq:scHeqnMcyl}) of our analogy, from the terms arising from the choice of coordinates or separation constants. As said before, since the spherical case for the $E$-field, equation~\eqref{eq:scHeqnEsph}, or the magnetic parts in cylindrical coordinates, equation~\eqref{eq:scHeqnMcyl}, will result in a prohibitively convoluted ODE determining the refractive index, we shall restrict ourselves to the $H$-field in the spherical case, equation~\eqref{eq:scHeqnMsph}, and the electric field in the cylindrical case, equation~\eqref{eq:scHeqnEcyl}. In these two cases, both equations are of the same form, as described above in equation~\eqref{eq:refindD}, with just minor changes in the precise form in which the corresponding separation constants appear: The relevant constant being $D=l(l+1)$ in the former, and $D=l^2-1/4$ in the latter case. Then the corresponding refractive index for creating an analogy for the radial part of the Teukolsky equation is given by
	\begin{equation}
	n(r,\omega) = \frac{c}{\omega}\sqrt{V(r,\omega) + \frac{D}{r^2}}.
	\end{equation}
	Written in such a way, it seems straightforward to give exact expressions for $n(r,\omega)$. This is true, however straightforward does not mean \enquote{manageable}: The (by far and wide) easiest way to achieve an exact expression involves a computer algebra system. We employed both {\sf Maple~2017} and {\sf Mathematica~11.2} for this purpose. In the limit $a\to 0$ corresponding to the Schwarzschild case the resulting refractive index is a simple special case and reads
	\begin{equation}
	n_\text{Schwarzschild}(r,\omega) = c\sqrt{\frac{(4D-27)M^2 + 2Mr(E-2D+11) + r^2\kl{r^2\omega^2 - E -4 +D}}{r^2 \omega^2 \kl{r-2M}^2}}.
	\end{equation}
	\begin{figure}
		\centering
		\includegraphics[width=.48\textwidth]{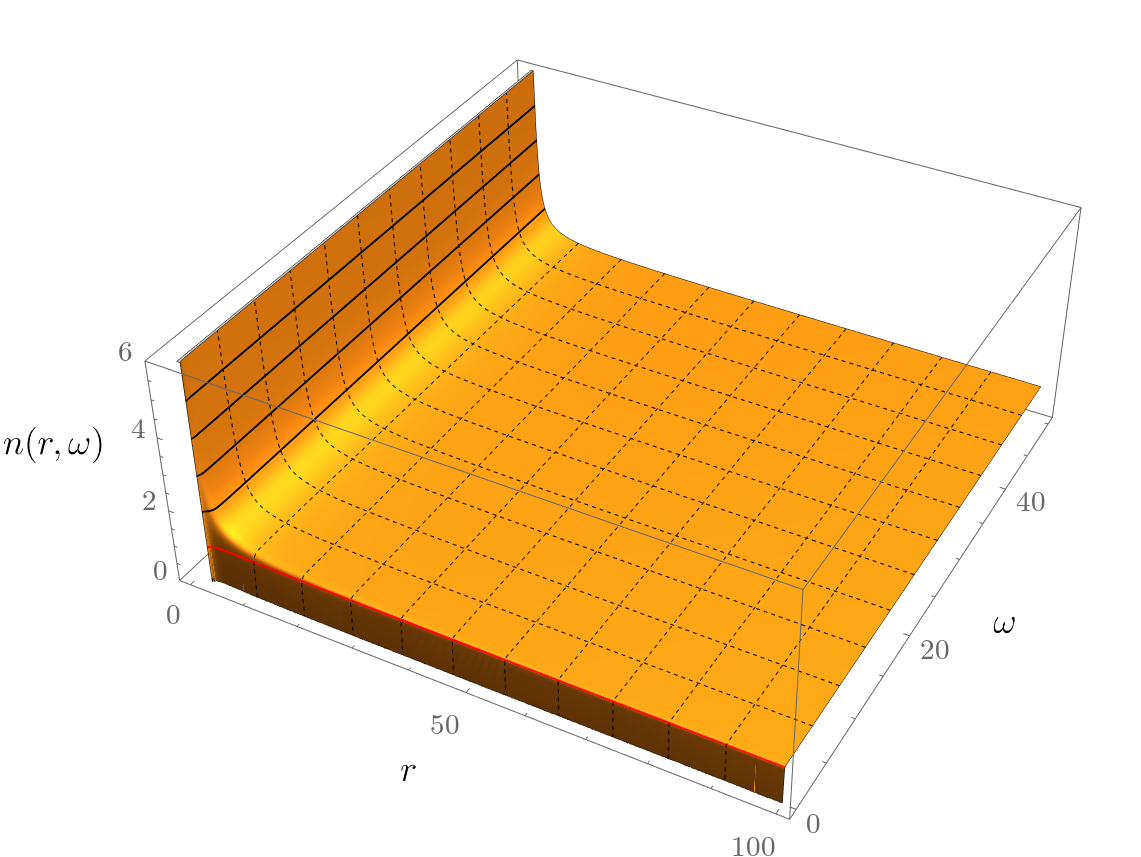}~\includegraphics[width=.48\textwidth]{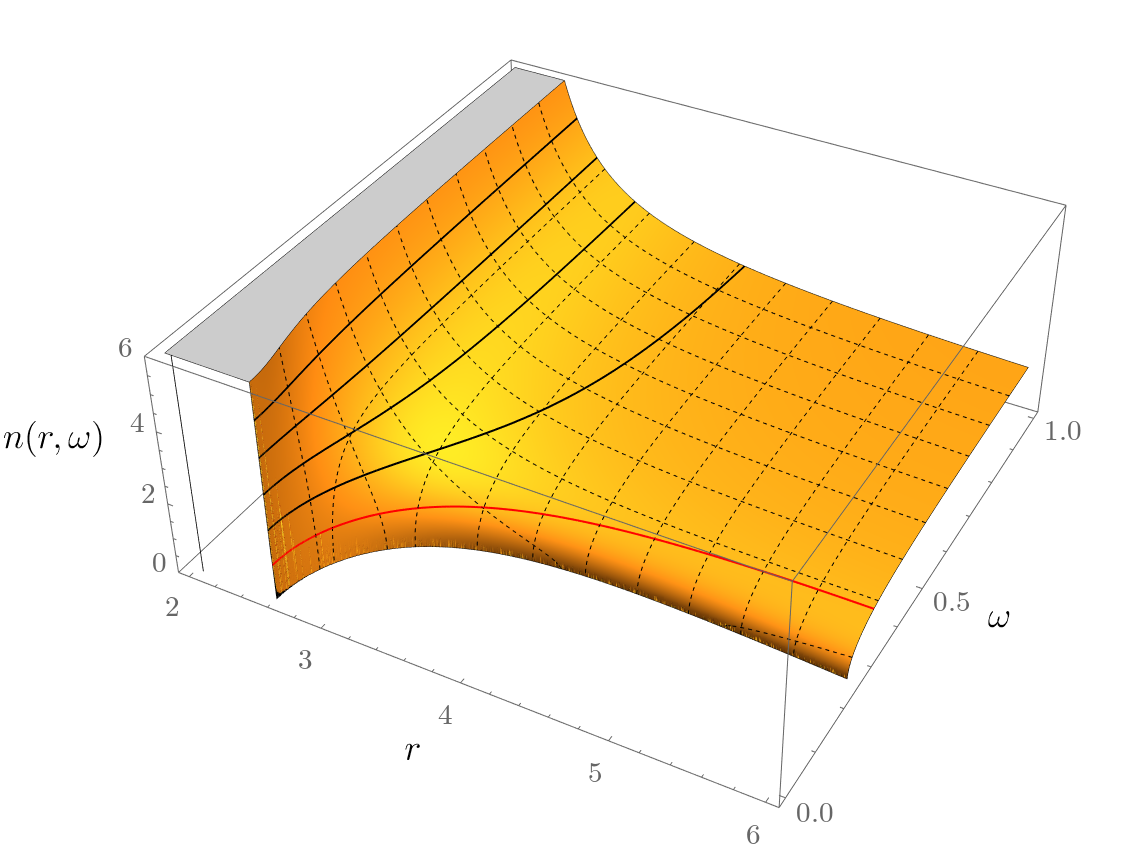}\\
		\includegraphics[width=.48\textwidth]{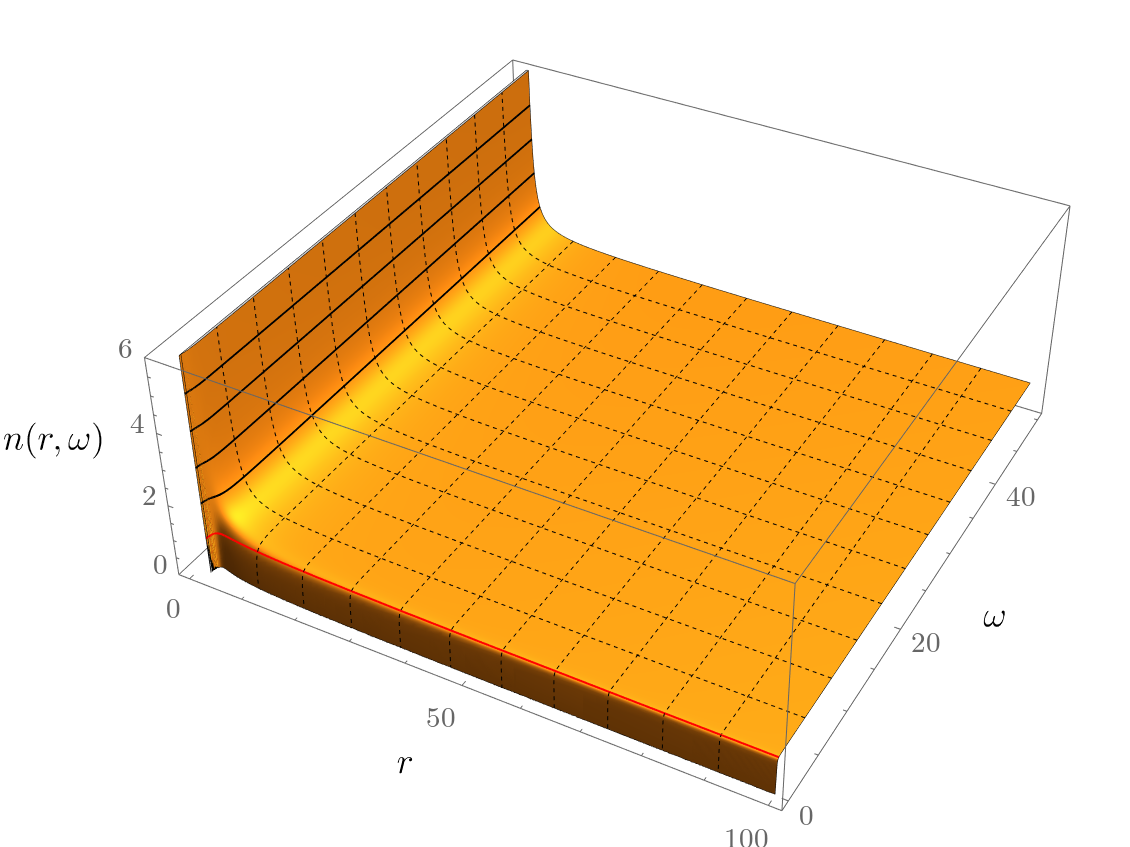}~\includegraphics[width=.48\textwidth]{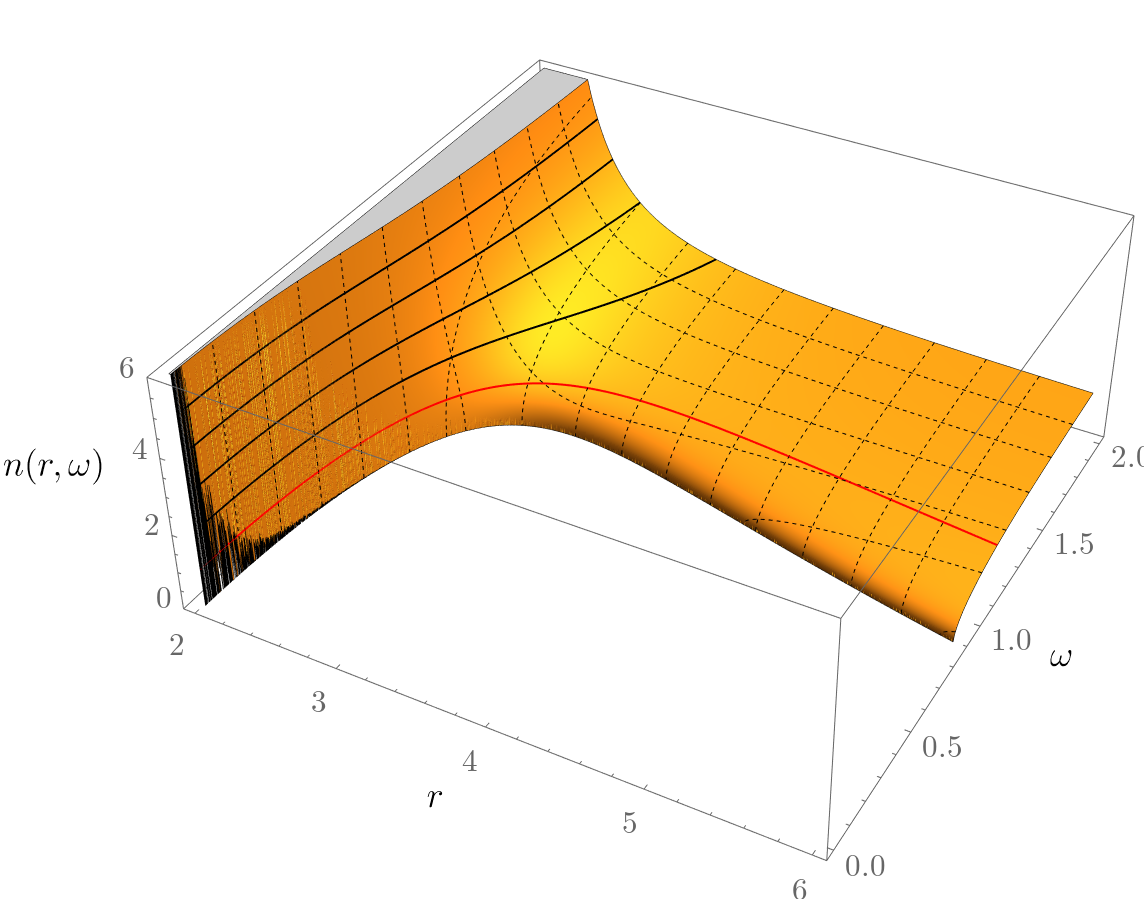}
		\caption{Indices of refraction for the Schwarzschild case, mass $M=1$, angular momentum $a=0$, and separation constants $D=0$. Dashed lines correspond to lines of constant $\omega$ or $r$. The red line is the contour of index of refraction equal to 1, that is, the vacuum value. Black lines are lines of constant $n\in\{2,3,4,5\}$, the cut-off is taken at $n=6$. \emph{In the left column:} A look at the larger picture. \emph{In the right column:} a zoom on smaller values of $r$ and $\omega$. \emph{Top row:} $\ell=1$ and $m=0$. \emph{In the bottom row:} $\ell=6$, $m=3$. The rugged structures visible in the bottom right plot for very small $r$ and $\omega$ are numerical artefacts.}
		\label{fig:refinSchwarzschild1}
	\end{figure}
	
	\begin{figure}
		\centering
		\includegraphics[width=.48\textwidth]{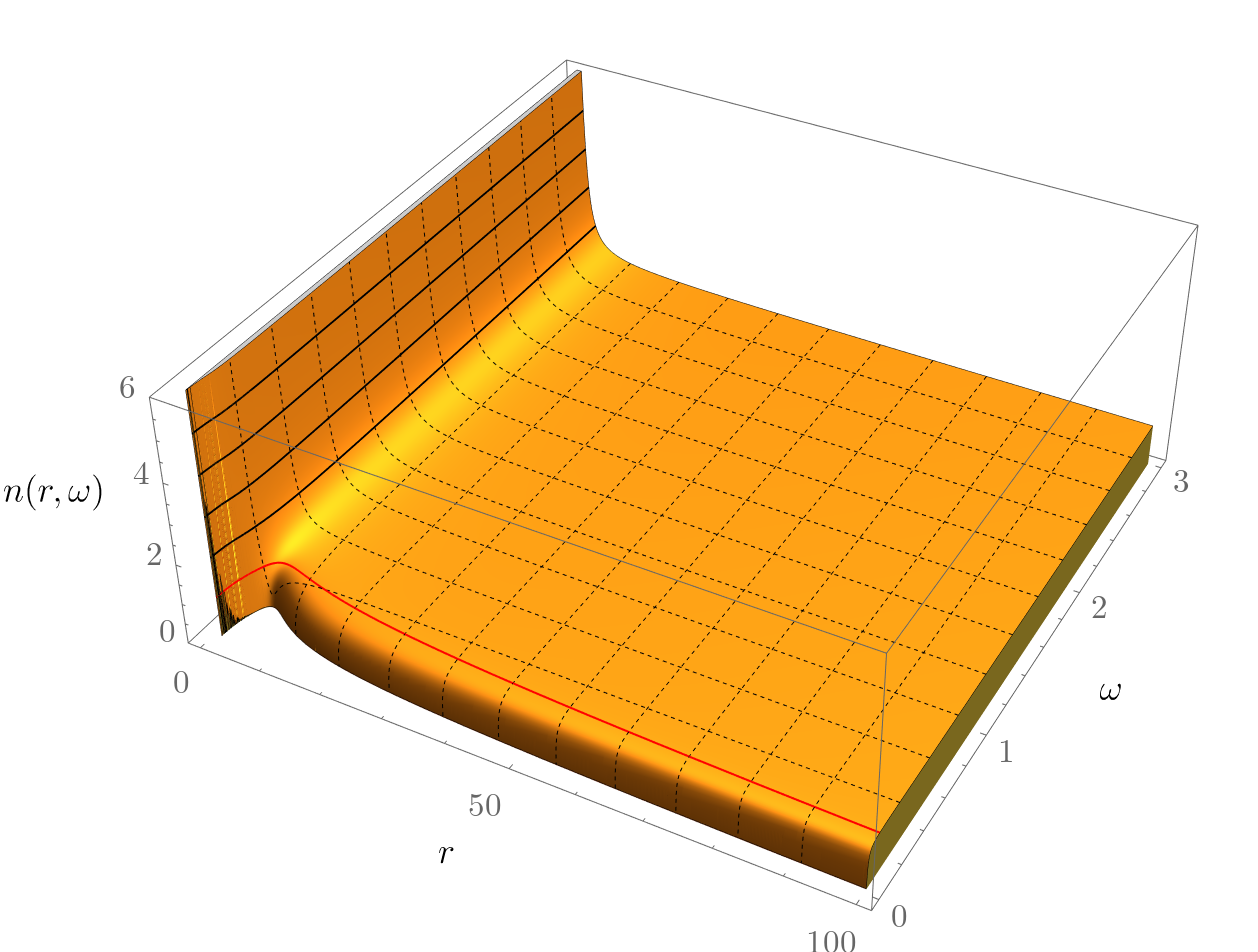}~\includegraphics[width=.48\textwidth]{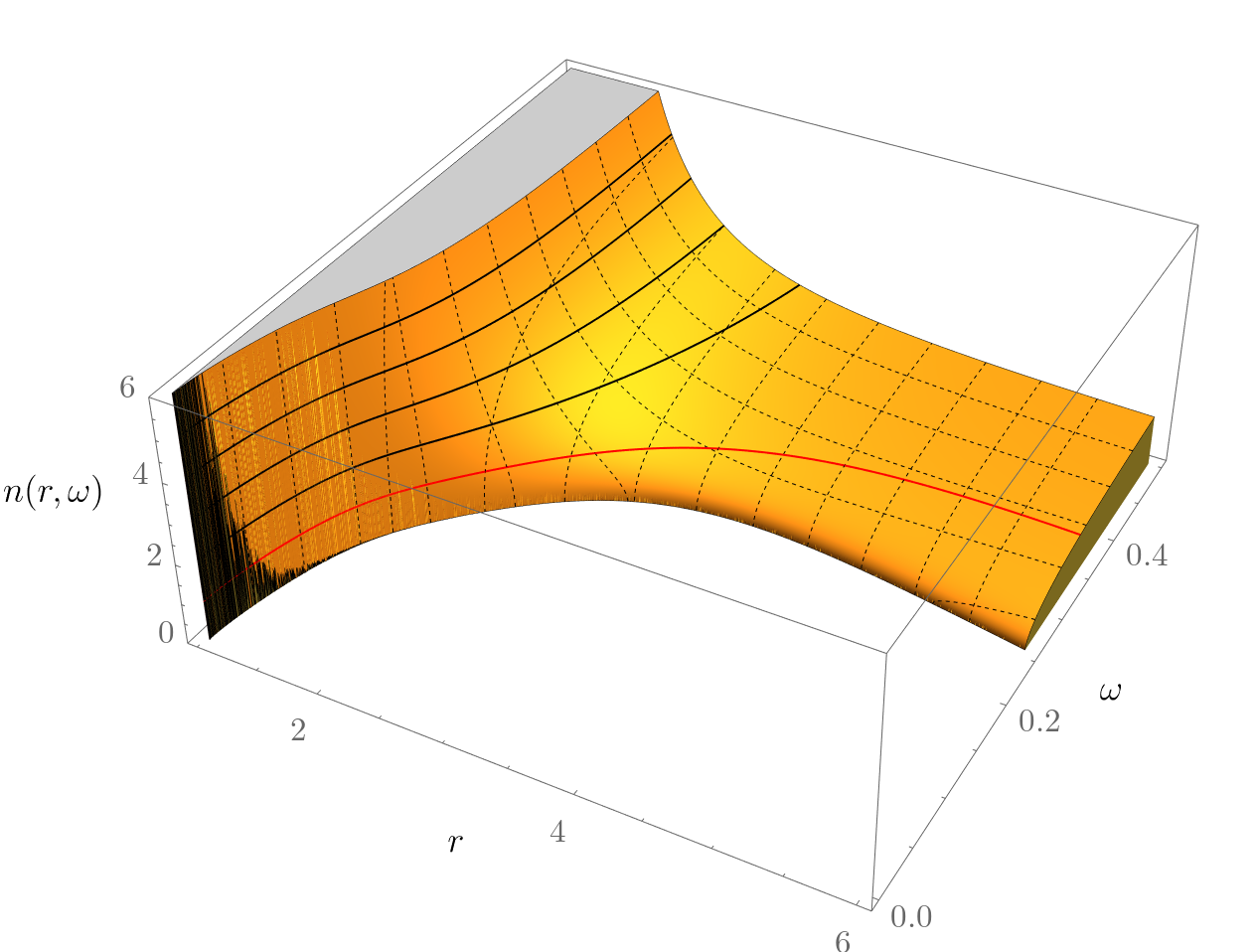}\\
		\includegraphics[width=.48\textwidth]{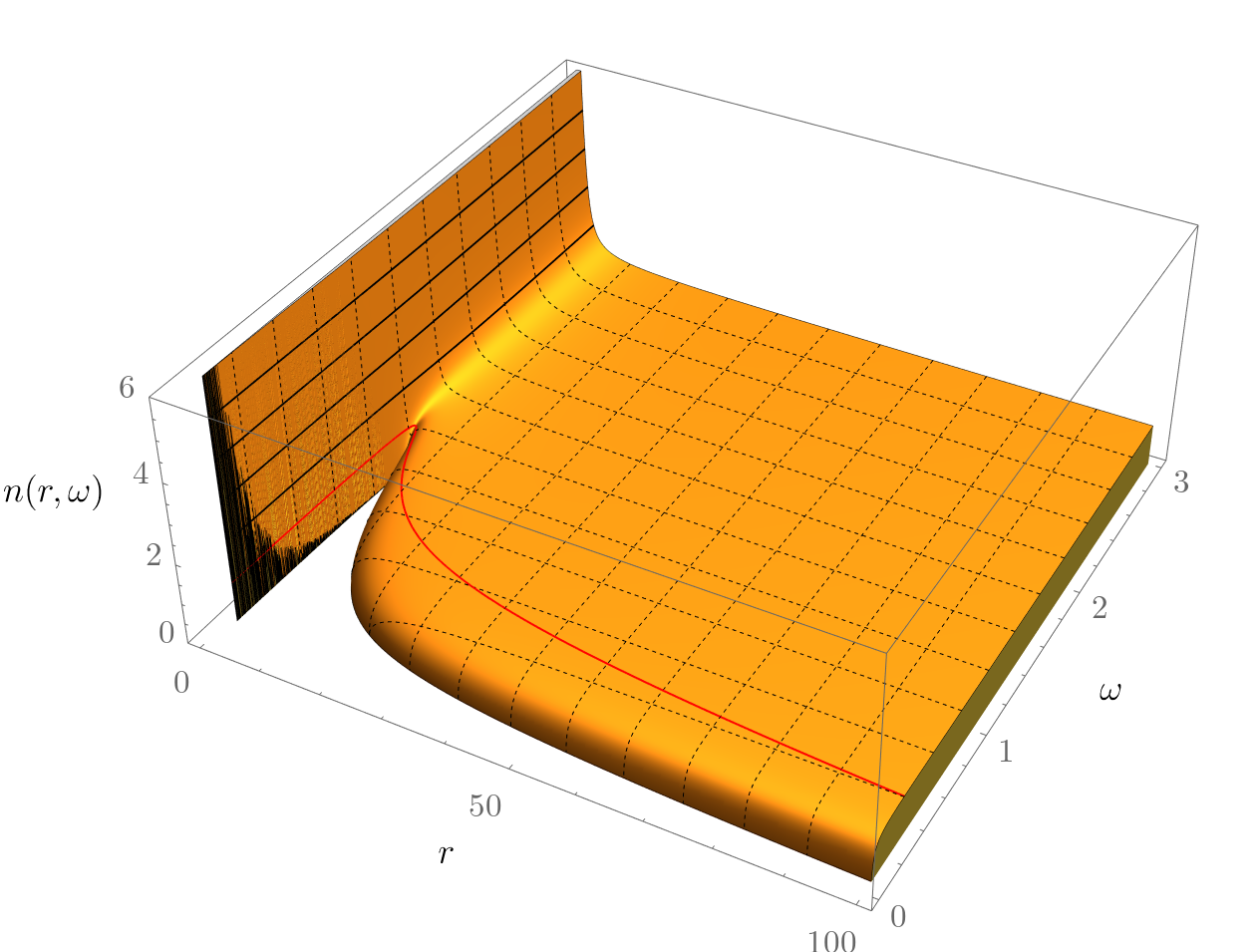}~\includegraphics[width=.48\textwidth]{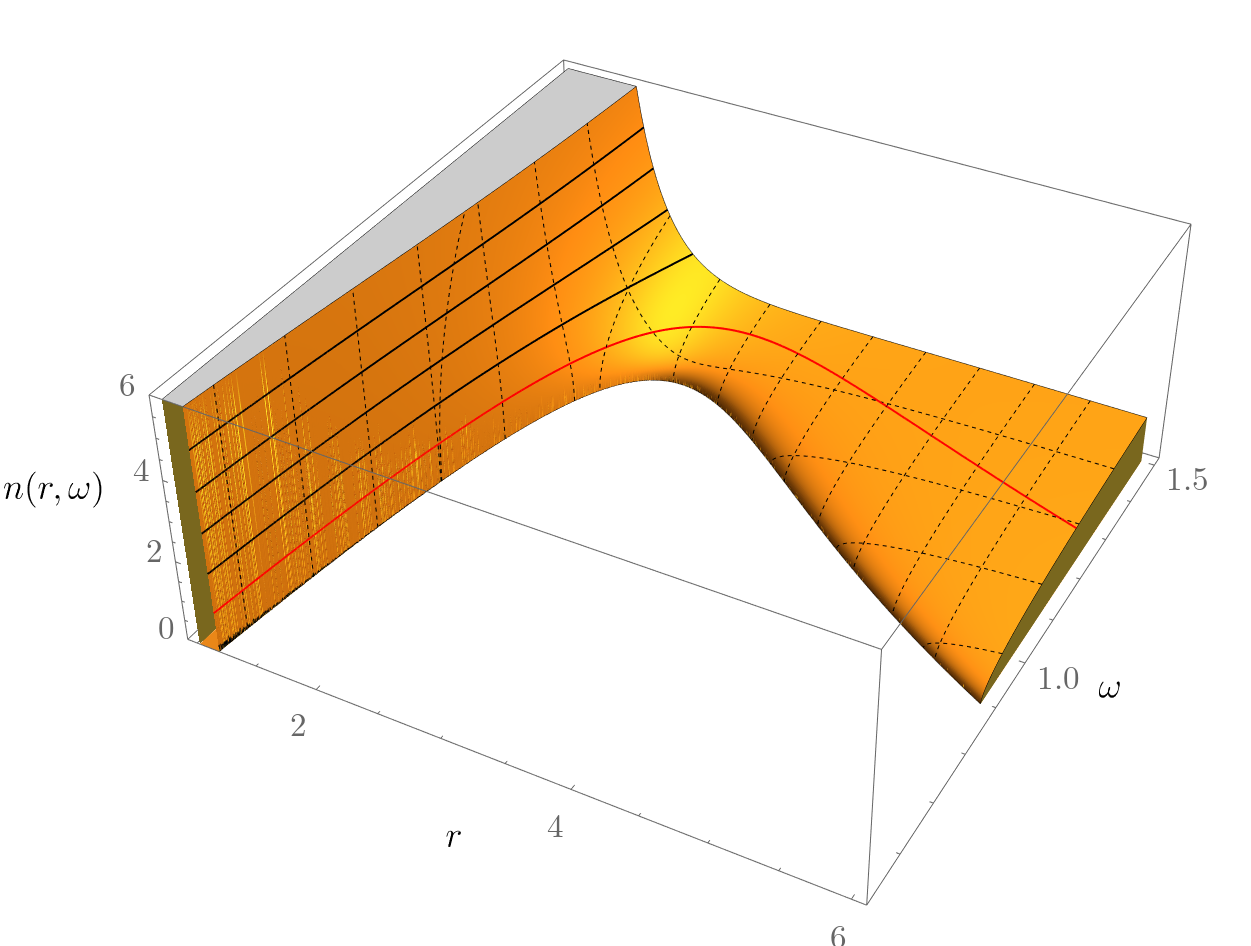}
		\caption{Indices of refraction for the Kerr case, mass $M=1$, angular momentum $a=0.99975$, and separation constants $D=0$. Dashed lines correspond to lines of constant $\omega$ or $r$. The red line is the contour of index of refraction equal to 1, that is, the vacuum value. Black lines are lines of constant $n\in\{2,3,4,5\}$, the cut-off is taken at $n=6$. \emph{In the left column:} A look at the larger picture. \emph{In the right column:} a zoom on smaller values of $r$ and $\omega$. \emph{Top row:} $\ell=1$ and $m=0$. \emph{In the bottom row:} $\ell=6$, $m=3$. As we used the data from \cite{TeukolskyPress74} for the separation constants of the Teukolsky equation, this time $\omega$ cannot exceed $\approx 3$. As the bottom right picture indicates, the fact that the diverging, real refractive index does not reach $\omega=0$ is a numerical artifact.}
		\label{fig:refinKerr3}
	\end{figure}
	
	\begin{figure}
		\includegraphics[width=.48\textwidth]{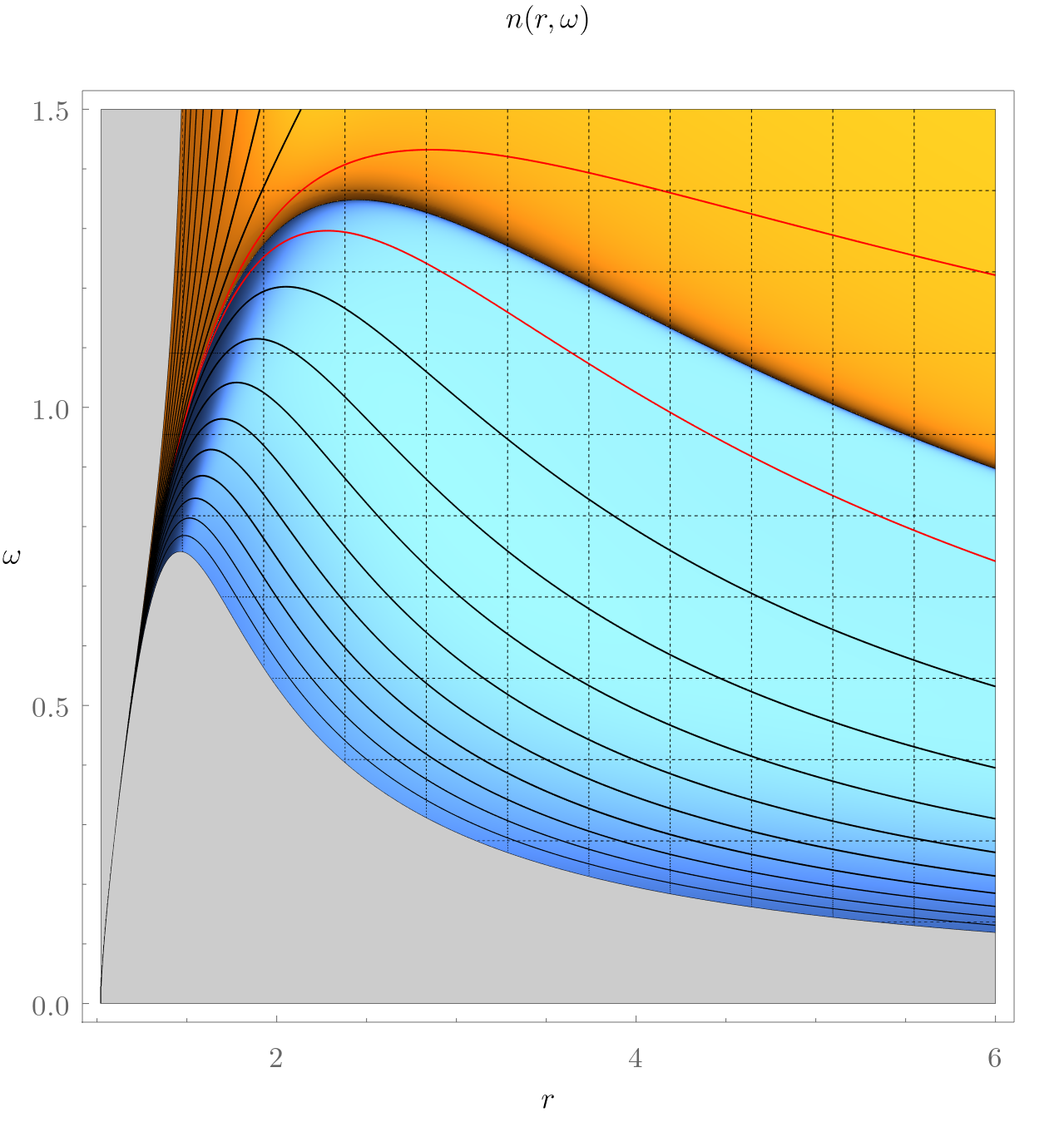}~\includegraphics[width=.48\textwidth]{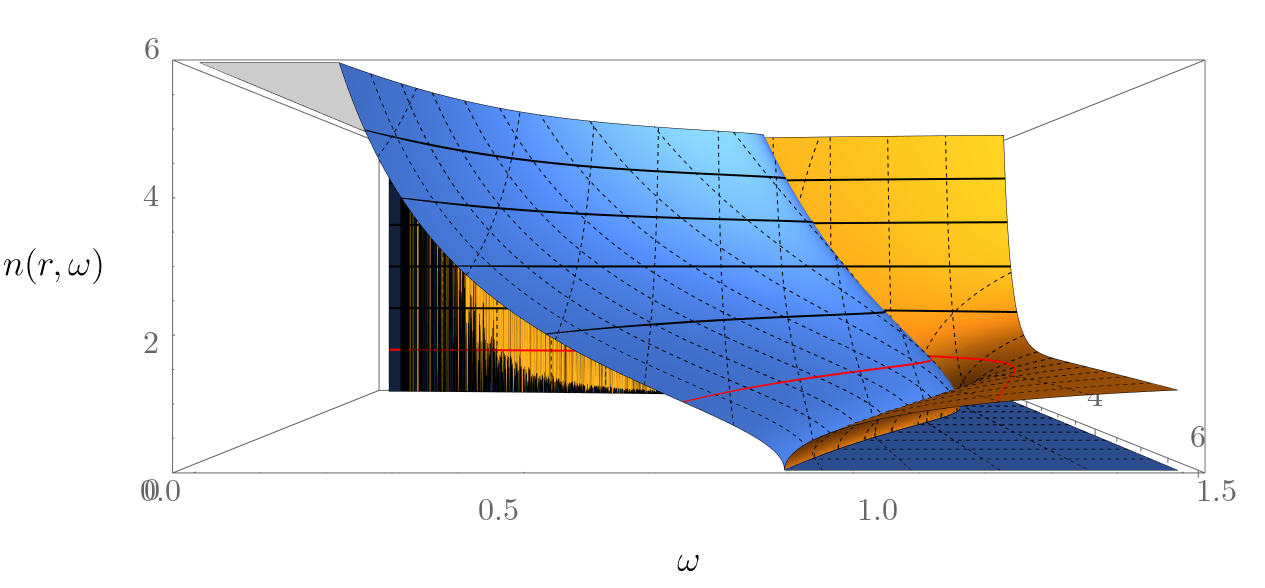}
		\caption{Real part (orange) and imaginary part (blue) of the Index of refraction for the Kerr case, mass $M=1$, angular momentum $a=0.99975$, and separation constants $D=0$. Dashed lines correspond to lines of constant $\omega$ or $r$. The red line is the contour of index of refraction equal to 1, that is, the vacuum value. Black lines are lines of constant $n\in\{2,3,4,5\}$, the cut-off is taken at $n=6$. \emph{Left:} View from above, showing contour lines. \emph{Right:} View from the front, demonstrating how the imaginary part vanishes in the ranges shown in the remaining figures.}
		\label{fig:refinReIm}
	\end{figure}
	
	In figures~\ref{fig:refinSchwarzschild1} and \ref{fig:refinKerr3} we can see the results for different values of modes --- of the separated Teukolsky equation (encapsulated in the separation constant $E$) --- and angular momentum, as described in their corresponding captions. The mass was set to be $M=1$. With the exception of figure~\ref{fig:refinReIm} we chose to plot only the real part of the refractive index for ease of viewing. While the refractive index does turn purely imaginary for low frequencies, for any value of $\omega$ there is a neighbourhood of the horizon in which it becomes real and positive again. However, any approach of the horizon leads to a divergence of the refractive index. The $\omega$ range of the fits is limited by the polynomial approximation used for $E_{\ell m}$. The exception is the Schwarzschild case, where the range of $\omega$ could be chosen freely, since $a=0$.
	
	Were we to change the separation constant $D$ of the Helmholtz equation, we can change the low-frequency behaviour if $D$ \enquote{outperforms} the contributions of $E_{\ell m}$. However, for any mass, any black hole rotation, any (non-zero) frequency, any mode, and any separation constant $D$, it holds that
	\begin{equation}
	\lim\limits_{r\to\infty} n(r,\omega) = 1. 
	\end{equation}
	The refractive indices hence pass the necessary consistency check: At spatial infinity, the refractive index has to reproduce the asymptotic flatness corresponding to the flat space Minkowski value of $n=1$.
	
	On the other hand, we saw that the refractive index diverges (to $+\infty$) when approaching the horizon. This is difficult to see in some of the figures, as the divergence close to $\omega\sim 0$ is very rapid and difficult to capture numerically. The numerical issues aside, this observation is in full agreement with the previously mentioned results of \cite{Bespoke1,Bespoke2} and \cite{ReznikRefIndexHorizons}: The optical properties have to diverge on the horizon/ergo-surface. In our concrete case, this divergence constitutes the biggest bane of the proposed analogy --- the refractive index of available materials rarely exceeds 4, usually only achieved by using meta-materials or near resonances (\emph{e.g.}, as done in \cite{XUVRefractiveOptics} using the Lorentz--Lorenz equation) and then only in small bandwidths. One additional example known to us is Lead(II) sulfide, see \cite{PbSRefInd} and \cite{HybridMaterials}, though for thin films or nano-particles. The experimental realisation hence remains far from trivial. But as we can see in the figures~\ref{fig:refinSchwarzschild1}---\ref{fig:refinReIm}, there the frequency-dependence is benevolent enough to allow for small bandwidths, at least.
	
	We have provided a new theoretical method to provide an analogy to radial propagation of electromagnetic waves in a Petrov type $D$ space-time using stratified refractive indices. In order to achieve this, we employed the separability of both the Helmholtz equation for the material in question, as well as the separability of the wave equation in these space-times. Since only an aspect of this wave equation --- radial propagation --- is mimicked, we opt to refrain from using the word \enquote{analogue} and rather refer to it as an \enquote{analogy}.
	
	\section{Conclusion}
	Let us summarise: We have seen that it is easily possible to find analogue space-time models that differ from the often-seen approach of rephrasing a given differential equation in terms of a Laplace--Beltrami equation involving an effective metric $\g$. Actually, it is worth to compare this with Gordon's 1923 paper. As one of the oldest analogue space-time models, it does not follow this line of reasoning if reexamined with our modern understanding. Rather, it is an \emph{algebraic} analogy that identifies $\g$. Similarly, while the second example, the \emph{analytic} analogy above, does arise out of differential equations, this time only parts of the full dynamics of a Laplace--Beltrami equation are recovered.
	
	We have also seen some usually overlooked issues with the algebraic model: Experimentalists trying to implement it will encounter additional terms in the Maxwell equations making the identification of the hidden analogue space-time vacuum electrodynamics difficult. This, however, can be turned on itself to also allow for analogue space-times whose connections are not the Levi-Civita connection of $\g$. Part of these additional terms are directly related to these differences of the two connections involved. This has the interesting consequence that this algebraic electromagnetic analogue provides access to more than just an analogue for general relativity: Also modified theories of gravity become available to the analogue space-time paradigm. This would, for once, allow a partly justified use of the more traditional name \enquote{analogue gravity}. Another part is the observation that coordinates matter again: Not the internal coordinates of the analogue space-time itself, but how these coordinates are mapped onto (labelled by) the laboratory coordinates.
	
	This strongly implies that coordinate artefacts, \enquote{cartographic distortions}, will be of an even more pernicious nature than in classical general relativity: Already general relativity itself had almost from its infancy to grapple with the question which effects are real and which are just due to a choice of coordinates --- examples being both the event horizon and the existence of gravitational waves (which was only settled once the sticky-bead-argument became accepted). The additional layer of coordinate transformations (from the laboratory space-time to the effective space-time or \emph{vice versa}) in the context of the algebraic analogue will only serve, sadly, to exacerbate this further. Underneath this tangle, however, will lie valuable lessons for effects of curved space-time physics, both classical and quantum mechanical in nature. We hope that methods of algebraic quantum field theory employed in the context of pre-metric electrodynamics, see \cite{PreMetricAQFT,PreMetricQEI}, might be able to shed more light on the precise nature of the Hawking effect in the algebraic analogue space-time.
	
	The analytic analogy luckily did not suffer from this: Here the interpretation of both coordinates and physical quantities is much more straightforward. The price one had to pay is a reduction in the dynamics that can be captured: Instead of a full partial differential equation, only one part of it is recovered. We imagine, though, that this analytic model might be easier to implement experimentally, as only the refractive index needs to be controlled, not permittivity, permeability, and magneto-electric tensor separately.
	
	In summary, both analogue models will provide exciting avenues for future experiments in the laboratory: The algebraic model could be achievable with bespoke meta-materials, should the properties described in \cite{Bespoke1,Bespoke2} be found to be in reach of current manufacturing technology. Even though the previous analysis shows that the interpretation will be non-trivial, it does not preclude the experiment as such. The analytic analogy might be realizable with gradient index (GRIN) methods. The interpretation of possible experiments is far simpler in this case.
	
	A commonality of both models is the divergence of optical properties when approaching the horizon (or, sometimes, the ergo-surface) of the analogue black hole. This might not preclude experimental insight into the quantum fields surrounding the analogue black hole, as renormalization calculations often do see significant effects away from the horizon. A demonstration of this can be found in \cite{BHQuantumAtmosphere}. We are looking forward to see these models further investigated by or together with experimentalists.
	
	\section*{Acknowledgements}
	Figure~\ref{fig:analoguecoords} is based on \url{http://www.texample.net/tikz/examples/cylinder-segment/} by Mathias Magdowski, published under (CC BY 2.5).\newline M.~Visser and S.~Schuster acknowledge financial support from the Marsden Fund administered by the Royal Society of New Zealand. S.~Schuster was also supported by a Victoria University of Wellington PhD Scholarship. The research presented herein is an extract of S.~Schuster's PhD thesis \cite{MyThesis} completed at the Victoria University of Wellington in December 2018. S. Schuster would like to thank Chris Fewster, Eleni Kontou, Eli Hawkins and Jörg Frauendiener for helpful discussions regarding the covariant macroscopic electrodynamics; Lorenz Drescher, Christian Gilfert, and Philipp Reiß-Jäger for brainstorming which materials could provide high refractive indices; and the anonymous referees for suggestions on how to improve the presentation.
	
\printbibliography
	
\end{document}